\documentclass{aa}  

\usepackage{graphicx}
\usepackage{txfonts}
\usepackage{lipsum}
\usepackage{subcaption}         
\usepackage{lscape}             
\usepackage{placeins}           
\usepackage{hyperref}
\usepackage{etoolbox}
\AtBeginDocument{%
  \nolinenumbers
  \let\linenumbers\relax
}

\begin{document}

   \title{Chandra and Multiwavelength Study of Pevatron Candiate HESS J1702--420A}

   \titlerunning{CXO Study of HESS J1702--420A}

%
%
%

   \author{Hui Yang\inst{1,2}\corrauth{hui.yang@cnrs.fr}        
        \and Yichao Lin\inst{2}\email{yichaolin@gwmail.gwu.edu}
        \and Jeremy Hare\inst{3,4,5}\email{harej10@gmail.com} \and Seth Gagnon\inst{2}\email{sethgagnon@gwmail.gwu.edu}
        \and Oleg Kargaltsev\inst{2}\email{kargaltsev@email.gwu.edu}
        }

   \institute{IRAP, UT/CNRS/CNES, 9 avenue du colonel Roche, 31028 Toulouse Cedex 04, France
   \and Department of Physics, The George Washington University, 725 21st Street Northwest, Washington, DC 20052, USA \and Astrophysics Science Division, NASA Goddard Space Flight Center, Greenbelt, MD 20771, USA \and Center for Research and Exploration in Space Science and Technology, NASA/GSFC, Greenbelt, Maryland 20771, USA \and The Catholic University of America, 620 Michigan Ave., N.E. Washington, DC 20064, USA}

   \date{Received September 30, 20XX}

\defcitealias{Giunti}{A21}
\defcitealias{2022A&A...667A.130G}{G22}
\newcommand{\Suzaku}{\emph{Suzaku} }

 
  \abstract
   {}
   {HESS~J1702--420A is a promising Galactic PeVatron candidate whose
  physical nature remains unknown. Previous \textit{XMM-Newton} observations revealed hard, apparently extended X-ray emission from XMMU~J170147.3--421407, offset from the center of the TeV source, suggesting a possible pulsar wind nebula (PWN) association. We aim to resolve its X-ray morphology, investigate compact-object candidates in the field, and assess their possible association with HESS~J1702--420A.} 
   {We obtained three new ACIS-I observations from the \textit{Chandra} X-ray observatory (CXO) with a total exposure of 50.4~ks. We performed imaging, source detection, spectral and variability analyses, and  multiwavelength (MW) machine-learning (ML) source classification. Neutron star (NS) candidates were further investigated using deep near-infrared and radio imaging.}
   {The \textit{Chandra} observations resolve XMMU~J170147.3--421407 into a faint point-like source,  Src11 (CXO J170144.5--421415), and asymmetric diffuse emission comprising eastern and western lobes. 
  The X-ray spectral, variability, and MW properties of Src11 and the diffuse emission are consistent with those commonly observed in pulsar/PWN systems, further supported by the ML classifications. 
  However, the asymmetric diffuse morphology is unlike a typical bow-shock/tail or jet/torus PWN, and the $\sim7'$ offset from HESS J1702--420A makes a firm physical association difficult to establish. A possible shell-like radio structure northwest of Src11 may indicate a more complex PWN/SNR environment. Other NS candidates are less consistent with a pulsar/PWN interpretation, making Src11 the most compelling candidate in the field.}
   {}


   \maketitle

\section{Introduction} \label{sec:intro} 

PeVatrons are Galactic sources responsible for accelerating cosmic rays up to 10$^{15}$ eV. 
However, despite several plausible associations \citep[i.e.,][]{2021Natur.594...33C,2021NatAs...5..460T}, no observation to date has established an unambiguous multi-wavelength (MW) counterpart directly identifying the accelerator responsible for PeV particle acceleration. 
Sensitive observations by the LHAASO, H.E.S.S., HAWC, and VERITAS TeV  $\gamma$-ray observatories have dramatically increased the number of known very high energy (VHE) $\gamma$-ray sources, which have been long recognized as tracers of  cosmic ray production sites, providing an insight into their origin. 
TeV observations have revealed an entirely new population of extended Galactic sources \citep{2005Sci...307.1938A,2009arXiv0907.0768C,2009arXiv0912.5355O},  many of which still have an unknown physical nature (see \cite{2013uean.book..359K,2020A&A...636A.113G,2018A&A...612A...2H} for a review). 

It is becoming increasingly apparent that relic (aged) pulsar-wind nebulae (PWNe; \citealt{2013uean.book..359K}) 
comprise the largest fraction of the population of  extended TeV sources found in the Galactic plane \citep{2013uean.book..359K,2020A&A...636A.113G,2018A&A...612A...2H,2010A&A...518A...8T,2010AIPC.1248...25K}.  
However, in many cases, the associations are based simply on positional coincidence, so there is always the possibility that a pulsar can be projected, by chance, within  the extent of the TeV source. 
The observed offsets between the pulsars and TeV sources and the asymmetries of the X-ray PWNe can be explained by either the pulsar proper motion or by the interaction with the reverse shock that had propagated through the interior of the nonhomogeneous supernova remnant (SNR) and reached one side of the PWN sooner than the other side \citep{2001ApJ...563..806B}. 
Often an association of a TeV source with a pulsar can be solidified if a PWN and its morphology are resolved in X-rays and/or radio. X-ray observations are particularly efficient  because PWNe can often be discovered and classified as such at  these energies, even in the absence of a known pulsar, due to their distinct properties.   
Choosing between different interpretations can be  challenging if only low-resolution  X-ray data are available, but the high angular resolution and low background of the Advanced CCD Imaging Spectrometer (ACIS; \citealt{2003SPIE.4851...28G}) onboard the {\sl Chandra} X-ray observatory (CXO) often allow one to confidently discriminate between the different interpretations based on X-ray images and spectra. Of course, there is always the possibility of discovering a completely new type of VHE source (especially in the densely populated central Galactic region) in which  case having high-resolution MW data is vital. 

\begin{figure*}
\begin{center}
\vspace{-0.2cm}
\begin{subfigure}[b]{0.49\textwidth}
        \centering
        \includegraphics[width=\textwidth]{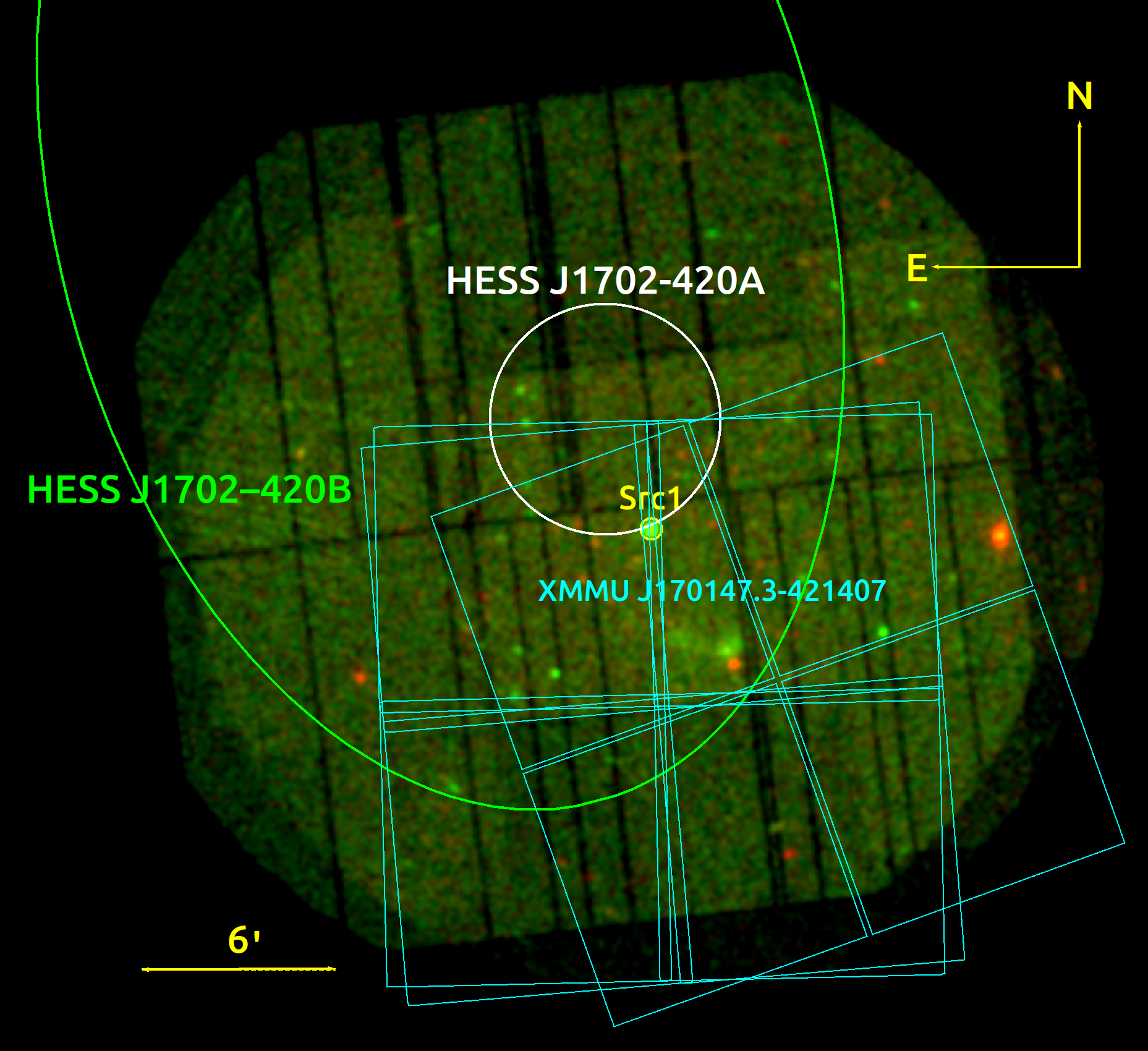}
    \end{subfigure}
    \hfill
    \begin{subfigure}[b]{0.49\textwidth}  
        \centering 
        \includegraphics[width=\textwidth]{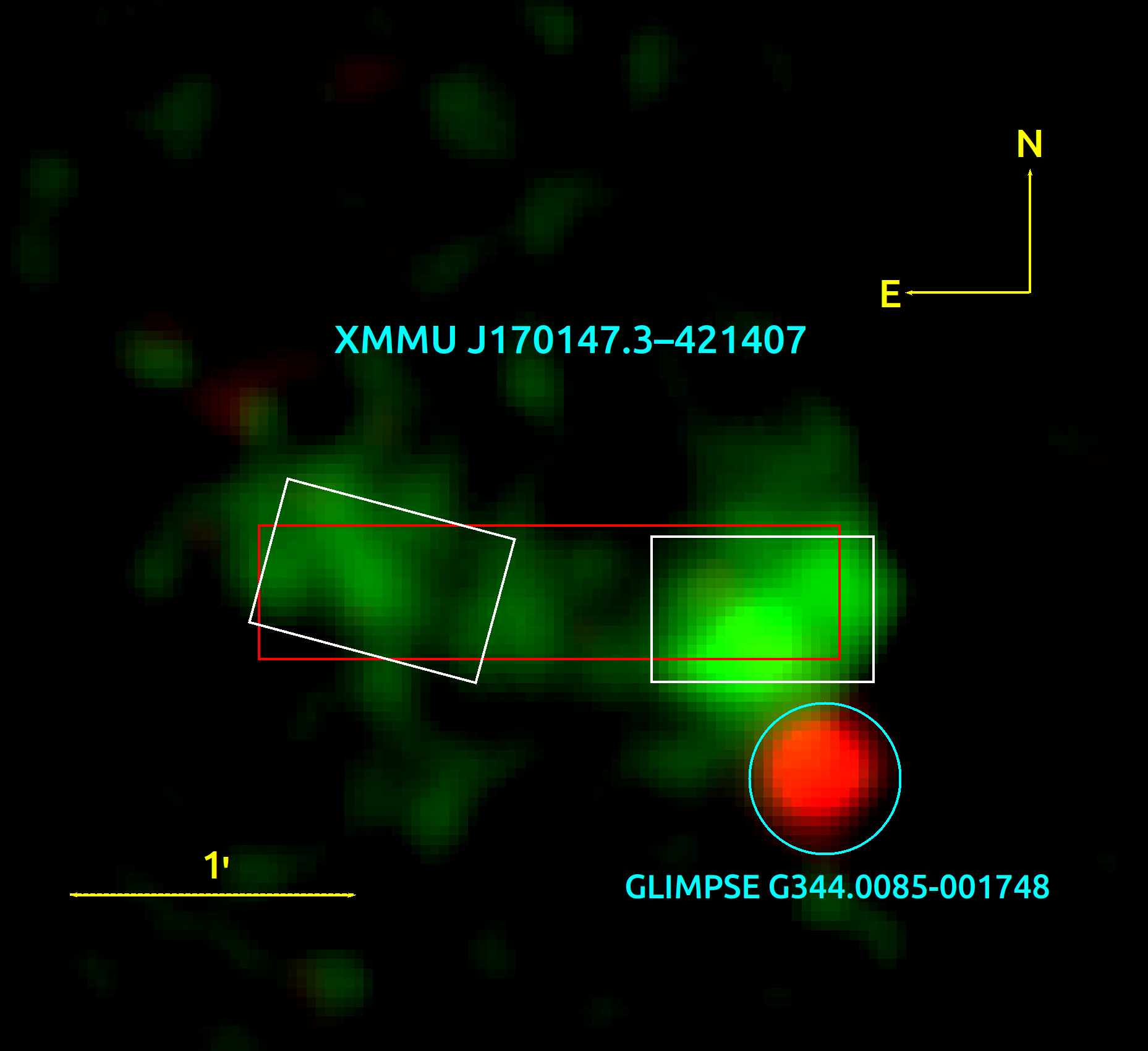}
    \end{subfigure}
\vspace{-0.5cm}
\end{center}
\caption{{\footnotesize The left panel shows the stacked two-color \textit{XMM-Newton} image of the HESS~J1702--420 field, with the 0.5--2~keV and 2--10~keV bands shown in red and green, respectively. The image was smoothed 
using a Gaussian kernel with a radius of 5 pixels and $\sigma=2.5$ pixels and is displayed with a square-root intensity scale. The $1\sigma$ uncertainty regions of HESS~J1702--420A and HESS~J1702--420B are overlaid in white and green, respectively. The FoV of the three \textit{Chandra} observations are shown in cyan, with the extended source XMMU~J170147.3--421407 located near the center of the combined coverage. \Suzaku Src~B (Src1 in this paper) is marked by the yellow circle. The right panel shows a zoomed-in view of the extended emission surrounding XMMU~J170147.3--421407. The soft X-ray source associated with GLIMPSE~G344.0085--001748 is marked by the cyan circle. The spectral extraction regions for the eastern and western lobes, as well as the larger region encompassing the overall extended emission used by \citetalias{2022A&A...667A.130G}, are shown by the white and red boxes, respectively.}}
\label{fig:XMM_img}
\vspace{-0.3cm}
\end{figure*}

\begin{figure*}
\begin{center}
\vspace{-0.2cm}
\begin{subfigure}[b]{0.49\textwidth}
        \centering
        \includegraphics[width=\textwidth]{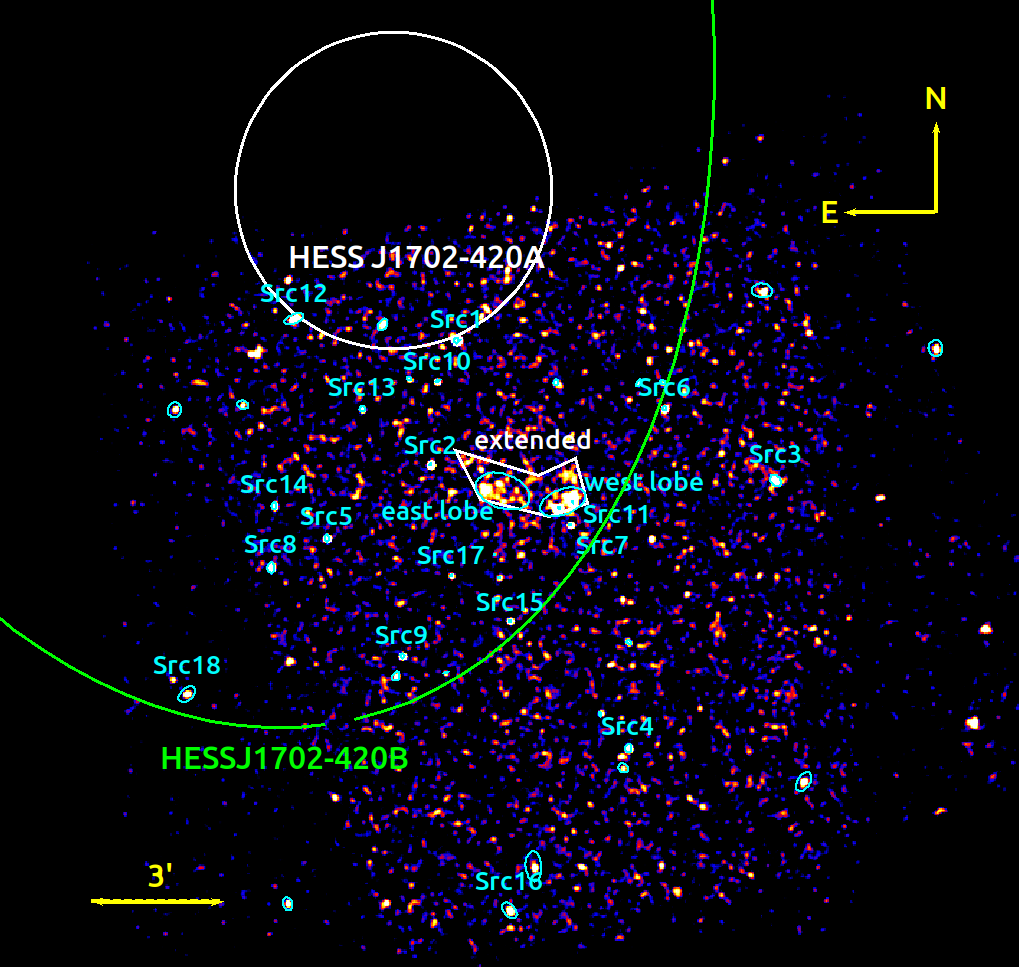}
    \end{subfigure}
    \hfill
    \begin{subfigure}[b]{0.49\textwidth}  
        \centering 
        \includegraphics[width=\textwidth]{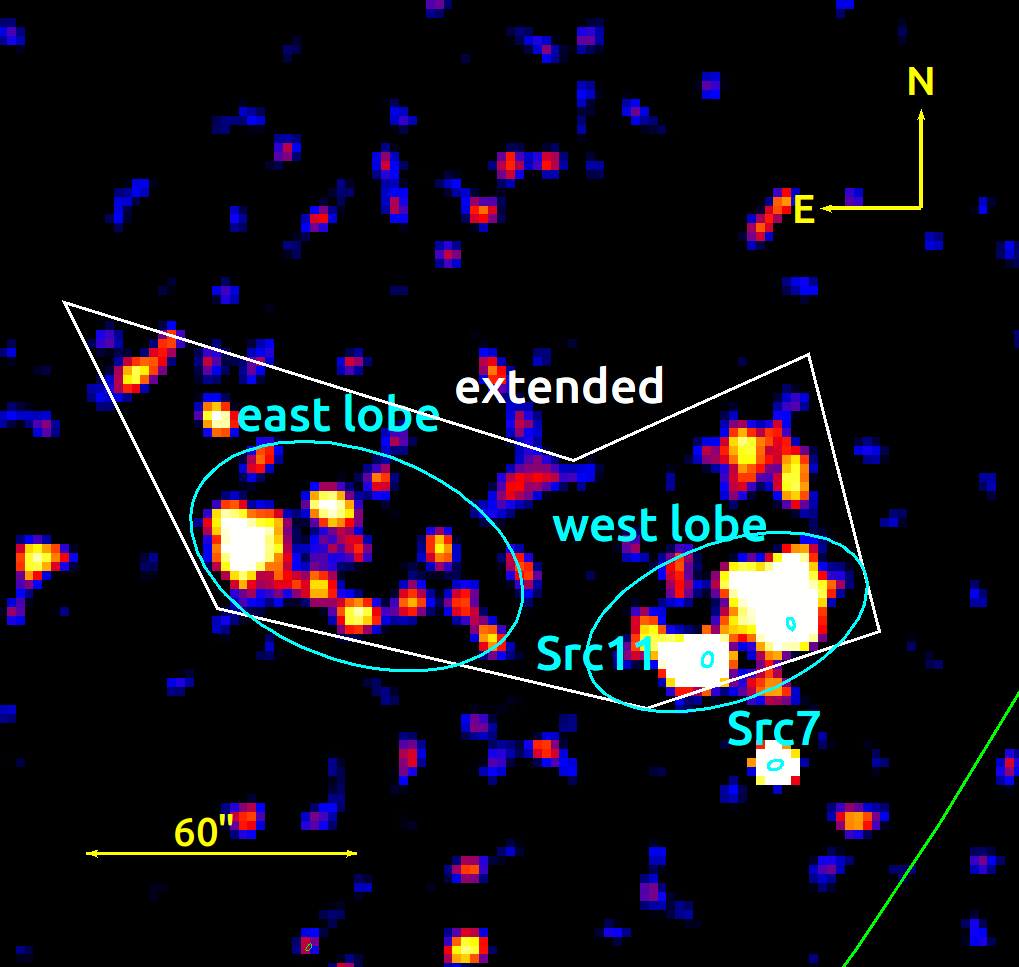}
    \end{subfigure}
\vspace{-0.5cm}
\end{center}
\caption{{\footnotesize The left panel shows the stacked \textit{Chandra} image in the 0.5--8~keV band, binned by a factor of 4 and smoothed using a Gaussian kernel with a radius of 3 pixels and $\sigma=1.5$ pixels, displayed with a logarithmic intensity scale. 
The cyan ellipses mark the X-ray sources detected by \texttt{wavdetect}; sources detected with a significance greater than $5\sigma$, which are analyzed in this work, are also labeled with their source names. The eastern and western lobes are indicated by cyan ellipses, while the region used to extract the spectrum of the combined extended emission is marked by the white polygon. The $1\sigma$ uncertainty regions of HESS~J1702--420A and HESS~J1702--420B are overlaid in white and green, respectively. The right panel shows a zoomed-in view of the extended emission surrounding XMMU~J170147.3--421407.}}
\label{fig:CXO_image}
\vspace{-0.3cm}
\end{figure*}

HESS J1702--420A was discovered in the analysis of deeper TeV observations of HESS J1702--420, an extreme particle accelerator and a promising PeVatron candidate found during the first H.E.S.S. Galactic plane survey \citep{Giunti}. 
The source has a flux of 2.08$\times10^{-13}$\,photons\,cm$^{-2}$\,s$^{-1}$ above 2 TeV and small size (the 1$\sigma$ radius is $(0.06\pm0.02_{\rm stat}\pm0.03_{\rm sys})^\circ$), overlapping with an elongated and much more extended companion TeV source HESS J1702--420B dominating the emission below a few tens of TeV (see the white circle for HESS J1702--420A and green ellipse for HESS J1702--420B in Figure \ref{fig:XMM_img} and \ref{fig:CXO_image}). 
An extremely hard power-law spectrum ($\Gamma=1.53$) 
and the lack of cut-off up to 100 TeV make HESS J1702--420A one of the hardest VHE $\gamma$-ray sources ever detected and a compelling candidate site for the presence of PeV cosmic rays \citep{Giunti}.  \cite{Giunti} also analyzed the archival Fermi-LAT data to constrain the source spectrum at $\gamma$-ray energies $>10$ GeV with a marginal detection of 4.3$\sigma$. Several scenarios have been proposed to explain the emission from HESS J1702--420. These include cosmic-ray (CR) diffusion from a nearby supernova remnant, SNR G344.7--0.1, and a pulsar, PSR J1702--412 (both $\sim0.5^\circ$ away from the centroid of the TeV emission; see, e.g., the discussion in \cite{Giunti}).  \cite{2023ApJ...955..147A} also proposed a model in which HESS J1702--420A is capable of accelerating protons to at least $\sim100$\,TeV while HESS J1702-420B is the more extended emission produced as the relativistic protons propagate through that dense molecular cloud. 

The source has been observed several times in X-rays in order to search for a potential MW association to the PeVatron candidate. Deep \Suzaku X-ray observations did not find a strong counterpart, reinforcing the ``dark accelerator" interpretation and supporting a hardronic PeVatron scenario \citep{fujinaga}. 
\citet{2024A&A...690A.142M} searched for diffuse X-ray emission associated with HESS~J1702--420 using \textit{eROSITA} data but found no extended counterpart, placing constraints on both hardronic and leptonic emission scenarios.  

A deep 72~ks \textit{XMM-Newton} observation (obsID=0884050101, see Table \ref{tab:cxo_obs} for details) of the HESS~J1702--420 field was analyzed by \cite{2022A&A...667A.130G} (hereby~\citetalias{2022A&A...667A.130G}). They found that \Suzaku src B (Src1 in our analysis, see Figure \ref{fig:XMM_img} and Figure \ref{fig:CXO_image}) is unlikely to be an isolated pulsar associated with HESS~J1702--420A, based on its strong absorption (\mbox{$N_\text{H}=(5.1\pm 1.1)\times 10^{22}\,\text{cm}^{-2}$}), Fe emission line(s) near 6.6~keV, long-term variability, and possible infrared (IR) counterpart (i.e., GLIMPSE G344.0901--00.1658), and instead suggested a magnetized cataclysmic variable (CV) interpretation. In contrast, XMMU J170147.3--421407 (See Figure \ref{fig:XMM_img} and Figure \ref{fig:CXO_image}), located $\approx7'$ from the center of HESS~J1702--420A, exhibited hard and highly absorbed X-ray emission with an apparently extended and asymmetric morphology. Its spectrum was well described by an absorbed power law with $\Gamma=1.6-1.8$ and $N_{\rm H}\approx3\times10^{22}$\,cm$^{-2}$), while the inferred absorption was broadly consistent with the cumulative hydrogen column along the line of sight at a distance of $\sim4.7$~kpc\footnote{Throughout this paper, we adopt $4.7$~kpc as the distance to XMMU J170147.3--421407 unless otherwise specified.}, albeit with large uncertainties.  Based on these properties, \citetalias{2022A&A...667A.130G} proposed XMMU J170147.3--421407 as a possible PWN powered by a high-velocity runaway pulsar, with the relic electron population potentially contributing to the VHE $\gamma$-ray emission from HESS~J1702--420A. 
However, the relatively low angular resolution and high background of the \textit{XMM-Newton} EPIC images prevented an unambiguous identification of a compact pulsar within the diffuse emission. High-angular-resolution \textit{Chandra} ACIS observations are therefore required to resolve the X-ray morphology and test the proposed PWN interpretation. 

The paper is structured as follows: we present the new \textit{Chandra} observations and their analysis in Sect. \ref{sec:CXO_Obs} whereas Sect. \ref{sec:discussion} provides a discussion of our results combined with the machine-learning classifications. In Sect. \ref{sec:conclusion} we will draw the conclusions.

\section{Chandra Observations}
\label{sec:CXO_Obs}

\subsection{Observations and Data Reduction}
\label{sec:obs_reduction}

\begin{table*}[ht!]
\caption{\label{tab:cxo_obs} Summary of  XMM and Chandra observations.}
\centering
\begin{tabular}{ccccc}
\hline\hline
ObsID & Observation Date & Exposure (ks) & Instrument & Mode \\
\hline
0884050101 & 2021 Sept. 26,27 & 80.4 & EPIC PN+MOS1,2 & Full Frame Mode + Medium Filter \\
\hline
30956 & 2025 June 03 & 18.66 & ACIS-I & Very Faint \\
30957 & 2025 June 08 & 9.96 & ACIS-I & Very Faint \\
28074  & 2025 Sept. 24 & 21.80 & ACIS-I & Very Faint \\
\hline
\end{tabular}
\end{table*}

We obtained three new observations of XMMU J170147.3--421407 from the Chandra X-ray Observatory (CXO) with a total scientific exposure of 50.42 ks (PI: H. Yang) (see Table \ref{tab:cxo_obs} for details). The data were collected using the ACIS-I chips. 
The software Chandra Interactive Analysis of Observations \citep[CIAO,][]{2006SPIE.6270E..1VF}, version 4.18, and the Chandra Calibration Data Base (CALDB), version 4.12.3, were used to reduce and analyze the data \footnote{\href{http://cxc.harvard.edu/ciao/}{http://cxc.harvard.edu/ciao/}}.  
The data from these observations were reprocessed using the {\tt chandra\_repro} CIAO tool.  

We performed an astrometry correction to the X-ray images with the details given in appendix \ref{sec:astrometry}. Then we run the CIAO Mexican-hat wavelet source detection routine {\tt wavdetect} \citep{Freeman2002} on the merged image with improved absolute astrometry to enable more accurate cross-matching to MW catalogs (discussed in Section \ref{sec:MUWCLASS}). 
We used an exposure-time-weighted average point-spread function (PSF) map and adhered to standard CIAO procedures\footnote{\url{http://cxc.harvard.edu/ciao/threads/}} to determine source fluxes, spectral properties, and variability. 

Each observation was processed using the CIAO tool \texttt{srcflux} to extract the fluxes in the soft (0.5--1.2 keV), medium (1.2--2.0 keV), hard (2.0--7.0 keV), and broad (0.5--7 keV) energy bands, based on the positions obtained by \texttt{wavdetect} on the merged image.  
We then used the CIAO tool \texttt{specextract} to extract spectra for each source, adopting the PSF 90\% encircled-count-fraction (ECF) aperture as the source regions and source-free circular regions with a radius of $1'$ on the same chip as the source for background extractions.  
Unless otherwise stated, all uncertainties throughout this paper are given at the 68\% confidence level, including the spectral parameters and the derived fluxes reported in Table \ref{tab:spectral_fit}. 
Below we only consider sources with significance $> 5$ which are sorted by their significance in Table \ref{tab:classification}.   

\subsection{Chandra Imaging}
\label{sec:cxo_image}

Figure \ref{fig:XMM_img} shows the the false color (0.5-2 keV -- red; 2-10 keV -- blue) XMM-Newton EPIC image of the HESS J1702--420 field, overlapping with the ACIS-I coverage from all three Chandra pointings (left panel) and the zoomed-in image of XMMU J170147.3--421407,  
using the same XMM-Newton observations from  \citetalias{2022A&A...667A.130G}. 

The left panel of Figure \ref{fig:CXO_image} shows the merged, astrometrically-corrected image from the entire ACIS-I in the 0.5-8 keV band. The confidently detected sources ($>5\sigma$) are labeled and spectral extraction regions for diffuse emission are also shown. The right panel offers an arcsecond-scale view of diffuse emission. With the new sub-arcsecond-resolution Chandra observations, the diffuse structure of XMMU J170147.3--421407 is clearly resolved into a point source, Src11 (CXO J170144.5--421415), embedded within the extended emission and spatially separated from Src7, which was difficult to achieve with the previous XMM-Newton observations presented by  \citetalias{2022A&A...667A.130G}. 

The most interesting sources from the part of HESS J1702--420 covered by ACIS-I are Src11 and the extended emission surrounding it. From both XMM and Chandra images in Figures ~\ref{fig:XMM_img} and \ref{fig:CXO_image}, the extended emission is strongly asymmetric and can be decomposed into two distinct substructures, hereafter referred to as the eastern and western lobes. The two lobes exhibit distinct morphologies and brightness distributions. In the right panel of Figure \ref{fig:CXO_image}, the western lobe is more compact and directly connected to Src11 through a continuous emission region. In contrast, the eastern lobe is broader and its emission appears to be discontinuous, with a noticeable decrease in surface brightness between the lobe and Src11. Although the western lobe is identified as a point source by \texttt{wavdetect}, its detection significance is low (1.7$\sigma$). Furthermore, its morphology is clearly extended rather than point-like. Therefore, we do not treat it as a separate point source, but instead consider it to be part of the extended emission. We used two elliptical regions and one polygon to extract spectra from the eastern and western lobes and the overall extended emission, respectively, as shown in Figure \ref{fig:CXO_image}. For the western lobe and the extended emission, the 90\% PSF ECF aperture of Src11 was excluded from the extraction regions. 

\subsection{Variability}
\label{sec:variability}

We investigated both the long-term and short-term variability of the detected point sources. To quantify variability between observations, we calculated the inter-observation variability probability parameter, $P_{\rm inter}$, by comparing the model-independent broadband fluxes derived with \texttt{srcflux} for the three \textit{Chandra} observations. Short-term variability within individual observations was assessed using Kuiper's test, and we defined the intra-observation variability parameter, $P_{\rm intra}$, as the highest probability of variability obtained among the three observations. We adopted the same procedures as those described in Section~2.2 of \cite{2022ApJ...941..104Y,2024ApJ...971..180Y}. Both parameters are included in Table \ref{tab:classification}. 

Among the 18 detected point sources, Src4 and Src17 show significant variability on both long and short time scales, with $P_{\rm inter}>0.95$ and $P_{\rm intra}>0.95$. 
Src2, Src3, Src6, Src7, Src10, Src11, and Src18 also show evidence for inter-observation variability, with $P_{\rm inter}>0.95$, among which Src2 and Src3 exhibit the strongest variability. For these sources we therefore allowed the normalization parameters to vary independently among the three observations during the spectral fitting. This enabled us to examine whether the model-dependent fluxes derived from the spectral fits exhibit variability consistent with that inferred from the model-independent broadband fluxes.

Src11 shows no evidence for short-term variability within the individual \textit{Chandra} observations, with $P_{\rm intra}=0.057$. For the long-term variability between observations, the measured energy fluxes are $(0.8^{+0.4}_{-0.3},\ 2.1^{+1.0}_{-0.7},\ 0.8^{+0.6}_{-0.4})\times10^{-14}\ {\rm erg\,cm^{-2}\,s^{-1}}$ in the 0.5--7 keV in the three observations, respectively. 
Although the model-independent variability analysis yields $P_{\rm inter}>0.95$, the fluxes derived from the spectral fits are statistically consistent  within their relatively large uncertainties. Thus, the current data provide no compelling evidence for significant long-term variability of Src11. Therefore, we tied the normalization in the spectral analysis for Src11 in Sec. \ref{sec:spectroscopy}. The \Suzaku and \textit{XMM-Newton} observations can not provide additional constraints on the variability of Src11 because their angular resolution is insufficient to resolve it from the surrounding diffuse emission. The absence of significant variability is consistent with an isolated pulsar interpretation than with an accreting compact object, although deeper observations are required to place stronger constraints on its variability properties. 

Src1, previously identified as Suzaku src~B by \cite{fujinaga}, also shows evidence for short-term variability, with $P_{\rm intra}=0.963$. However, no significant long-term variability is found for this source based on the comparison of its fluxes among the three \textit{Chandra} observations. 
Given the possible compact-object nature of Src1, we additionally searched for periodic X-ray pulsations using the \textit{XMM-Newton}/EPIC-pn data. Following the procedure outlined in \cite{2023ApJ...951...80H}, we performed a periodicity  search on the barycenter corrected photon arrival times using a merged event list (i.e., MOS1+2 and PN). We searched over a frequency range of 5$\times10^{-5}$ Hz to $\sim0.2$ Hz , with the high frequency end determined by the time resolution of the MOS1+2 instruments (2.6 s).  We used the $Z_n^2$ test \citep{1983A&A...128..245B} and  adopted a 5$\sigma$ trials-corrected detection threshold. No statistically significant periodic signal was detected. Thus, within the sensitivity and temporal-frequency range of the present EPIC data, we find no evidence for coherent X-ray pulsations from Src1.

\subsection{Chandra Spectroscopy}
\label{sec:spectroscopy}

The spectra of the sources and diffuse emission were jointly fitted using all three Chandra observations. 
The spectral fits were performed using XSPEC (v12.14.1; \citealt{1996ASPC..101...17A}) with an absorbed power-law (PL) model, \texttt{tbabs*powerlaw}. The \texttt{tbabs} model accounts for interstellar photoelectric absorption using the default vern photoelectric absorption cross-sections \citep{1996ApJ...465..487V} and the angr solar abundances \citep{1989GeCoA..53..197A}. 
The spectra are grouped to at least one photon per bin for fitting and are fitted with the WSTAT statistic \citep{1979ApJ...228..939C}. The energy range for the fits was restricted to 0.5-7 keV. 
Sources with $P_{\rm inter}>0.95$ were fitted with their normalization parameters free between the three observations.  
The results of the spectral fitting are summarized in Table~\ref{tab:spectral_fit} with the plots available at  \href{https://huiyang-astro.github.io/HESS_J1702-420/}{https://huiyang-astro.github.io/HESS\_J1702-420/}.

The spectra from the hard diffuse emission from the western and eastern lobes were fitted separately first. The two regions show somewhat different spectral parameters: the western lobe, which is more compact and closer to Src11, has a hard spectrum with $\Gamma=0.9^{+0.7}_{-0.6}$ and poorly constrained $N_{\rm H}=1.4^{+1.4}_{-1.1}\times10^{22}\,\mathrm{cm^{-2}}$, while the more extended and more distant eastern lobe has a softer spectrum ($\Gamma=2.1^{+0.8}_{-0.7}$) with $N_{\rm H}=3^{+2}_{-1}\times10^{22}\,\mathrm{cm^{-2}}$. The two measurements of $N_{\rm H}$ are consistent within their relatively large uncertainties, and both individual fits provide acceptable fit statistics. 
Although the spectral slopes are marginally different and may indicate spectral evolution, the uncertainties are large enough to justify a joint fit to the extended emission from the western and eastern lobes. 
We therefore also performed a joint fit to the combined extended emission, including both the western and eastern lobes. 
The fit gives intermediate $\Gamma\approx1.8\pm0.5$ and $N_{\rm H}\approx2.2^{+0.9}_{-0.8}\times10^{22}\,\mathrm{cm^{-2}}$, with a comparable fit statistic. 
Fixing $N_{\rm H}$ at the best-fit value of $2.2\times10^{22}\,\mathrm{cm^{-2}}$, the photon indices of the western and eastern lobes become more comparable: $\Gamma=1.3\pm0.3$ and $1.7\pm0.3$, respectively. 
The combined diffuse emission has an observed 0.5-7 keV flux of $(1.5\pm0.1)\times10^{-13}\mathrm{erg\,cm^{-2}\,s^{-1}}$ with the western and eastern lobes contributing $(6.2^{+0.6}_{-0.5}\times10^{-14})$ and $(5.6^{+0.7}_{-0.6}\times10^{-14})\,\mathrm{erg\,cm^{-2}\,s^{-1}}$, respectively. 

Src11, which may be physically related to the nearby diffuse emission assuming a pulsar/PWN scenario (see Section~\ref{sec:pulsar_PWN}), is very faint. 
The source only has 17 photons with $<1$ expected to come from the background.  
Not surprisingly, its spectrum is poorly constrained when both $N_{\rm H}$ and $\Gamma$ are allowed to vary, giving $N_{\rm H}\lesssim2.6\times10^{22}\,\mathrm{cm^{-2}}$ and $\Gamma=0.4^{+1.7}_{-0.9}$. 
As we don't find convincing evidence for variability across observations (see Sec. \ref{sec:variability}),  
we tied the normalization between the three observations while keeping $N_{\rm H}=2.2\times10^{22}\,\mathrm{cm^{-2}}$ fixed. In this final fit, the photon index is $\Gamma=1.4\pm1.0$, with an observed flux of $(0.9^{+0.3}_{-0.2})\times10^{-14}\,\mathrm{erg\,cm^{-2}\,s^{-1}}$. 

\begin{table*}
\caption{Spectral fitting results.} \label{tab:spectral_fit}
\centering
\begin{tabular}{llccccc}
\hline
Source$^a$ & CXO Name$^b$ & Counts$^c$ & $W/{\rm d.o.f.}^d$ & $N_{\rm H}$ & $\Gamma$ & $F_{\rm X}$ (0.5--7 keV)$^e$ \\
 & & & & ($10^{22}\ {\rm cm}^{-2}$) & & ($10^{-14}{\rm erg\,cm^{-2}\,s^{-1}}$) \\
\hline
Src1 & J170157.2--421027 & 236 & 155.42/214 & $4.7^{+1.6}_{-1.4}$ & $1.5\pm0.5$ & $18\pm1$ \\
Src2 & J170200.4--421316 & 73 & 68.50/57 & $3.2^{+1.4}_{-1.3}$ & $2.5^{+0.8}_{-0.7}$ & $(0.9^{+0.5}_{-0.4},0.4^{+0.5}_{-0.3},6\pm1)$ \\
Src3 & J170117.9--421338 & 93 & 77.34/80 & $0^{+5.6}_{-0}$ & $-0.8^{+1.1}_{-0.4}$ & $(11^{+2}_{-1},6^{+2}_{-1},5\pm1)$ \\
Src4 & J170136.0--421944 & 47 & 32.82/38 & $1.9^{+2.0}_{-1.7}$ & $3.8^{+1.5}_{-1.4}$ & $(1.7^{+0.5}_{-0.4},3\pm1,1.3^{+0.4}_{-0.3})$ \\
Src5 & J170213.1--421456 & 34 & 27.75/27 & $8.5^{+10.7}_{-7.7}$ & $1.8^{+2.8}_{-2.4}$ & $2.1\pm0.4$ \\
Src6 & J170131.6--421200 & 33 & 33.64/24 & $2.2^{+4.0}_{-2.2}$ & $1.8^{+1.4}_{-1.2}$ & $(2.3^{+0.7}_{-0.6},1.6^{+0.7}_{-0.5},1.0^{+0.4}_{-0.3})$ \\
Src7 & J170143.2--421439 & 19 & 7.89/11 & $2.4^{+1.9}_{-2.3}$ & $7.7^{+?}_{3.9}$ & $(0.2^{+0.3}_{-0.2},1.3^{+0.5}_{-0.4},0.6\pm0.2)$ \\
Src8 & J170220.0--421537 & 39 & 35.29/33 & $1.7^{+4.3}_{-1.7}$ & $0.1^{+1.3}_{-1.0}$ & $2.3\pm0.4$ \\
Src9 & J170203.9--421737 & 26 & 26.43/20 & $4.2^{+3.7}_{-2.8}$ & $2.9^{+1.8}_{-1.5}$ & $1.0\pm0.2$ \\
Src10 & J170159.5--421122 & 20 & 5.66/11 & $0^{+1.4}_{-0}$ & $2.7^{+1.9}_{-0.8}$ & $(2.8^{+0.9}_{-0.7},3\pm2,0.9^{+0.8}_{-0.5})$ \\
Src11 & J170144.5--421415 & 17 & 8.67/9 & $0^{+2.6}_{-0}$ & $0.4^{+1.7}_{-0.9}$ & $(0.8^{+0.4}_{-0.3},2.1^{+1.0}_{-0.7},0.8^{+0.6}_{-0.4})$ \\ 
Src12 & J170217.2--420957 & 38 & 33.11/32 & $17^{+17}_{-12}$ & $4^{+4}_{-3}$ & $1.9^{+0.4}_{-0.3}$ \\
Src13 & J170208.8--421200 & 16 & 4.62/10 & $0.4^{+1.6}_{-0.4}$ & $3.5^{+2.5}_{-1.6}$ & $0.7\pm0.2$ \\
Src14 & J170219.6--421412 & 21 & 18.39/16 & $0^{+8.2}_{-0}$ & $1.5^{+2.0}_{-1.1}$ & $2.2^{+0.6}_{-0.5}$ \\
Src15 & J170150.6--421650 & 12 & 5.82/6 & $0^{+10}_{-0}$ & $-1.5^{+2.6}_{-?}$ & $1.4^{+0.5}_{-0.4}$   \\
Src16 & J170150.7--422325 & 32 & 20.79/26 & $16^{+28}_{-15}$ & $1^{+5}_{-3}$ &$2.5^{+0.6}_{-0.5}$ \\
Src17 & J170157.8--421547 & 11 & 4.08/4 & $?$ & $0.7^{+1.4}_{-1.5}$ & $(5^{+6}_{-3},2.2^{+0.9}_{-0.7})$ \\
Src18 & J170230.4--421829 & 24 & 20.62/17 & $0^{+1.7}_{-0}$ & $-2.1^{+1.1}_{0.9}$ & $(2.6^{+0.9}_{-0.8},3\pm1)$ \\
extended & & 635 & 418.35/483 & $2.2^{+0.9}_{-0.8}$ & $1.8\pm0.5$ & $15\pm1$ \\
eastern lobe & &207 & 181.44/184 & $3^{+2}_{-1}$ & $2.1^{+0.8}_{-0.7}$ & $5.5^{+0.7}_{-0.6}$ \\
western lobe & & 185 & 165.12/169 & $1.4^{+1.4}_{-1.1}$ & $0.9^{+0.7}_{-0.6}$ & $6.5\pm0.6$ \\
\hline
Src11 & J170144.5--421415 & 17 & 12.01/12 & 2.2 (fixed) & $1.4\pm1.0$ & $0.9^{+0.3}_{-0.2}$ \\
extended & & 635 & 418.35/484 & 2.2 (fixed) & $1.8\pm0.2$ & $15\pm1$ \\
eastern lobe & & 207 & 181.80/185 & 2.2 (fixed) & $1.7\pm0.3$ & $5.6^{+0.7}_{-0.6}$ \\
western lobe & & 185 & 165.52/170 & 2.2 (fixed) & $1.3\pm0.3$ & $6.2^{+0.6}_{-0.5}$ \\
\hline
\end{tabular}
\tablefoot{
The second part of the table presents the results of refitting Src11 and the extended emission with $N_{\rm H}$ fixed and the normalizations tied. \\
$^a$ Source name used in this paper. \\
$^b$ CXO name in the form "Jhhmmss.s+ddmmss", based on the ICRS position, with an accuracy of 0.1s in RA and 1.0s in Dec. \\
$^c$ Combined total counts from the three observations. \\
$^d$ W-statistic per degree of freedom. \\
$^e$ Broadband fluxes. For sources with $P_{\rm inter}>0.95$, multiple values are reported for the individual observations. Some sources have only two values because they are not covered by the third observation. 
}
\end{table*}

\section{Discussion}
\label{sec:discussion}

\subsection{Pulsar and PWN candidates}
\label{sec:pulsar_PWN}

\begin{figure}
\begin{center}
\vspace{-0.2cm}
\includegraphics[width=0.49\textwidth]{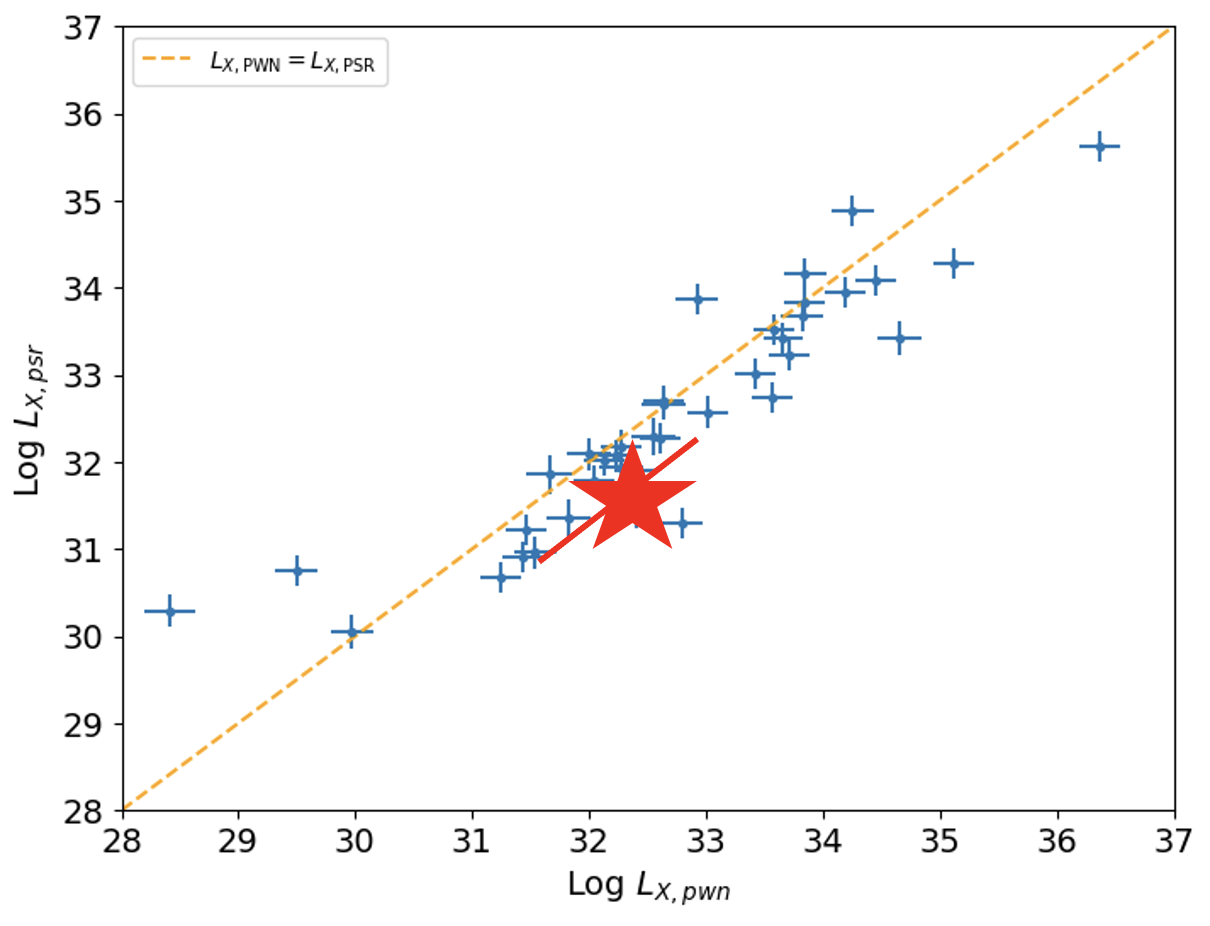}
\vspace{-0.5cm}
\includegraphics[width=0.49\textwidth]{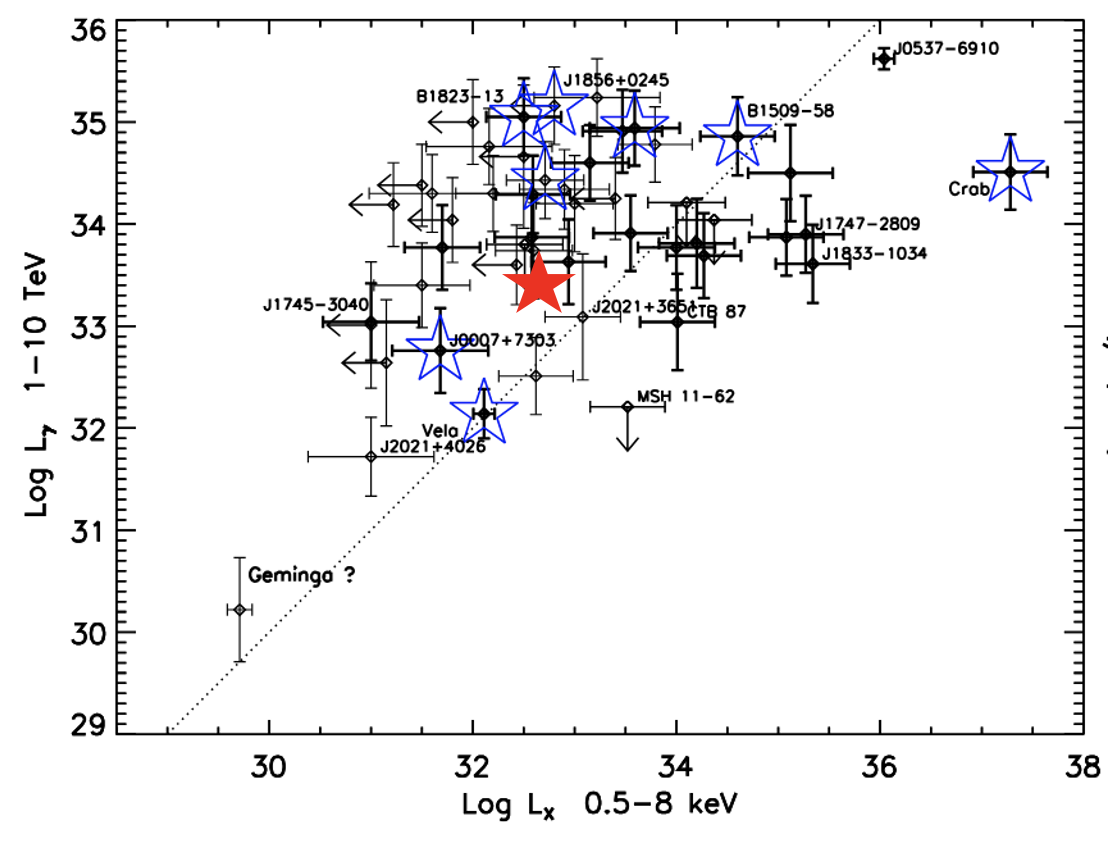}
\end{center}
\caption{{\footnotesize Top: Luminosities of the pulsars and their respective nebulae in the 0.5--8 keV. The flux for the extended region is used to calculate the PWN luminosity in this band. The red line shows the distance range of 2 -- 10 kpc. The orange dashed line indicates $L_{\rm X,pwn}=L_{\rm X,psr}$.
Bottom: Adapted from Figure 4 in \cite{2013uean.book..359K}}. The dotted line indicates $L_\gamma=L_X$. The putative pulsar/PWN system associated with Src11 and HESS J1702--420A is shown by the red star.} 
\label{fig:Luminosities}
\vspace{-0.3cm}
\end{figure}

There are two main pieces of evidence that the Src11 + western-lobe + eastern-lobe system may constitute a pulsar and its associated PWN: namely, its morphology and spectral properties. The eastern lobe is clearly not point-like, while both components within the western lobe also appear to be spatially extended. Additionally, the spectra of Src11, the eastern lobe, western lobe, and the entire extended region are all adequately described by an absorbed power-law model with consistent values of $N_{\rm H}$. 
If these sources are located at a similar distance, their spatial proximity and consistent absorption make a chance association between physically unrelated extended sources unlikely. 
The hard spectra of Src11, the western lobe, and the eastern lobe are also consistent with the range of non-thermal X-ray photon indices observed in pulsar/PWN systems \citep[$\Gamma\sim0.6-2.4$, i.e., ][]{2003ApJ...591..361G}. 
There is a known correlation between the luminosity of pulsars and the luminosity of their respective PWNe. Using a distance range of 2--10 kpc and the X-ray unabsorbed fluxes from the Src11 and extended regions respectively (see Table~\ref{tab:spectral_fit}) this system follows the typical behavior of other pulsar/PWN systems (see top panel of Figure~\ref{fig:Luminosities}). 
The X-ray/$\gamma$-ray luminosity of the extended diffuse emission of PWNe are also known to exhibit some correlation. Assuming a distance of $4.7$~kpc, the X-ray luminosity derived from the flux of the extended region, together with the $\gamma$-ray luminosity of HESS J1702--420A derived from its spectral model, falls within the range occupied by the known PWN population (see bottom panel of Figure~\ref{fig:Luminosities}, adapted from \citealt{2013uean.book..359K}). 

The morphology of the eastern and western lobes, however, does not readily resemble that of a conventional tail associated with a supersonic PWN. In such a scenario, the putative pulsar could currently power a bow-shock PWN, while the offset VHE emission of HESS J1702--420A could represent a "relic PWN" under a leptonic scenario, analogous to systems such as PSR J1740+1000 \citep{2026ApJ...997..341G}. 
Assuming a distance of 4.7 kpc and a typical pulsar velocity of $v_{\rm p}=200$~km~s$^{-1}$, a pulsar would require approximately $47$~kyr to traverse the $\sim7'$ angular separation between HESS J1702--420A and XMMU J170147.3--421407. Such an age would be consistent with a middle-aged pulsar, which may have escaped its host SNR and subsequently developed a bow-shock PWN. However, the morphology of the observed lobes does not strongly support this interpretation. The western lobe does not point back toward HESS J1702--420A, while the eastern lobe points only approximately in this direction and becomes brighter with increasing distance from Src11, contrary to the behavior generally expected for a simple pulsar tail. Moreover, the brightness profiles of the lobes are not suggestive of opposite sides of a bow-shock apex, which would generally be expected to exhibit approximate symmetry about the pulsar and decreasing surface brightness with increasing distance from the apex.

Alternatively, if the putative pulsar has not yet escaped its host SNR, it could power a subsonic PWN, with the western and eastern lobes potentially representing a jet and counterjet, respectively. The increase in brightness along the lobes could, in principle, result from interaction with the SNR reverse shock or from ram-pressure collimation of the outflow, which could enhance the local magnetic field and consequently the synchrotron emissivity \citep[e.g.,][]{2026ApJ..1008..139G}. 
The best-fit photon indices of the two lobes are $\Gamma_{\rm west}=1.3\pm0.3$ and $\Gamma_{\rm east}=1.7\pm0.3$, respectively. Although the eastern lobe appears softer, the two photon indices are statistically consistent within their uncertainties. 
Therefore, the spectral difference does not provide strong evidence against a jet/counterjet interpretation, and the measured photon indices are broadly consistent with those reported by \citetalias{2022A&A...667A.130G}. 
If the two structures constitute a jet and counterjet originating from the same particle population, broadly similar spectral properties might be expected.

We note that similarly disjoint/complex morphologies of non-thermal X-ray emission have been observed in other pulsar/PWN systems, which may represent an amorphous or strongly distorted PWN whose morphology does not follow the more commonly observed bow-shock/tail or jet--torus configurations. For example, deep \textit{Chandra} observations of the Vela-like pulsar PSR J1809--1917 revealed a compact PWN together with more extended, arcminute-scale ``hook''-like structure.  Moreover, the pulsar is offset by $\sim8'$ from the brightest part of the nearby TeV source HESS J1809--193, with the extended X-ray PWN aligned toward the TeV emission, supporting their association \citep[e.g.,][]{2007ApJ...670..655K,2018ApJ...861....5K}. 

\begin{figure}
\begin{center}
\vspace{-0.2cm}
\includegraphics[width=0.49\textwidth]{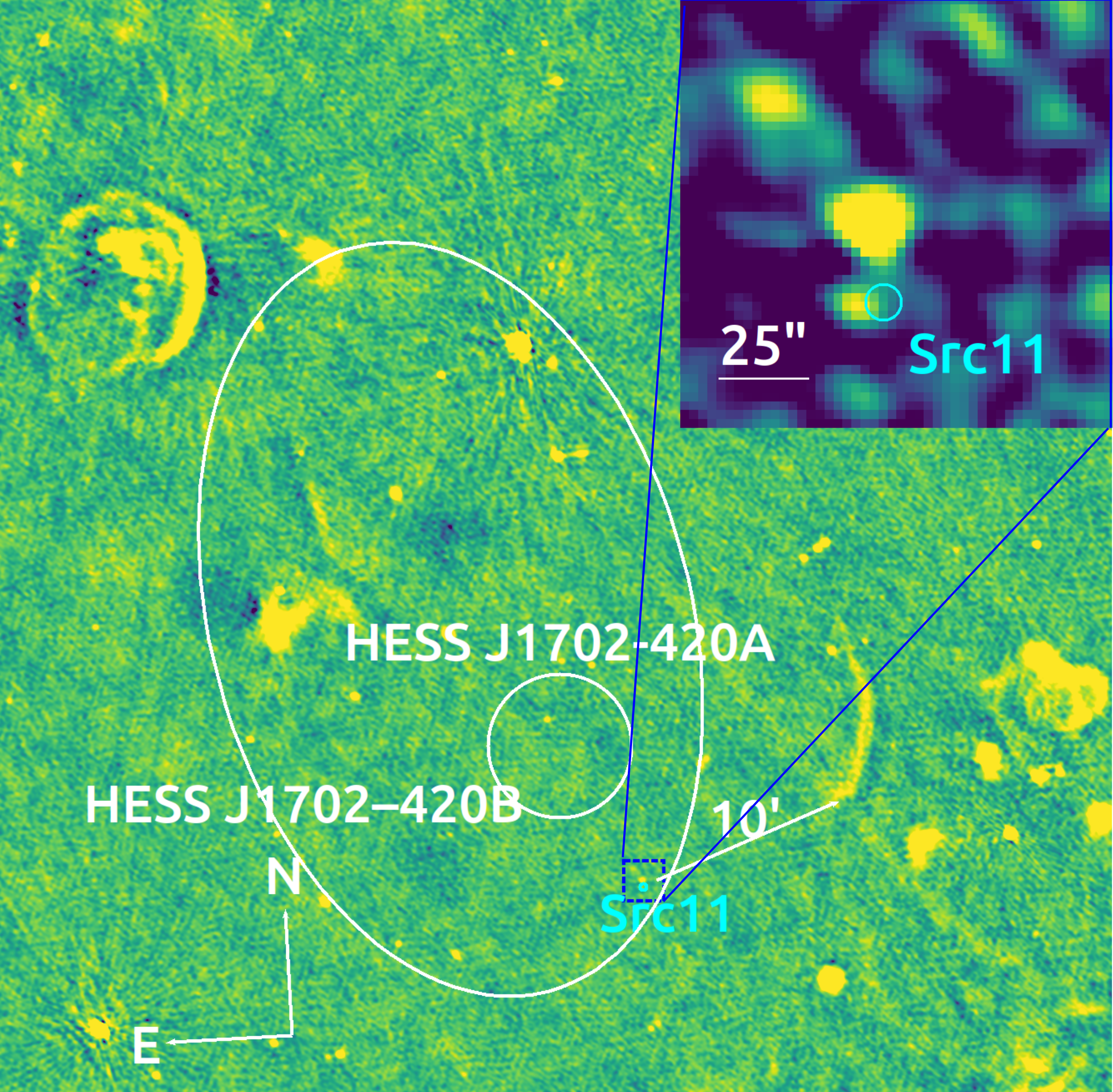}
\vspace{-0.5cm}
\end{center}
\caption{The RACS low-frequency intensity image. The 1-$\sigma$ error ellipses of HESS J1702--420A and HESS J1702--420B are overlaid in white, and Src11 position is overlaid in cyan. The 10$'$ white arrow points from Src11 to the south end of the shell-like radio structure. The inset shows the zoom-in image of Src11. } 
\label{fig:RACS}
\vspace{-0.3cm}
\end{figure}

We also note a shell-like feature in the Rapid ASKAP Continuum Survey \citep[RACS;][]{2023PASA...40...34D} low-frequency intensity image located approximately $10'$ northwest of Src11 (see the arrow from Src11 pointing toward the southern end of the radio structure in Figure~\ref{fig:RACS}), which could potentially represent part of an SNR shell. However, the curvature of the arc suggests a relatively small shell size, making it difficult to account for the $\sim7'$ offset between Src11 and the centre of HESS~J1702--420A if the two were physically associated. A faint, possibly extended radio source is also detected approximately $12''$ north of Src11 (see the zoom-in image in Figure~\ref{fig:RACS}). This source may represent an unrelated background radio-loud active galactic nucleus (AGN) with extended jets, although no X-ray counterpart is detected. Alternatively, it may constitute part of the larger shell-like radio structure. In addition, Src11 itself has no convincing optical or NIR counterpart, which makes an AGN interpretation involving X-ray-emitting jets less likely. The nature and relationship of these radio features therefore remain uncertain. If the shell-like structure is confirmed to be part of an SNR physically associated with the HESS~J1702--420A region, particle acceleration within the SNR, or interactions between accelerated hadrons and nearby dense material, could potentially contribute to the observed TeV emission.

\subsection{MUWCLASS results}
\label{sec:MUWCLASS} 

\begin{figure*}
    \centering
    \begin{subfigure}[b]{0.45\textwidth}
        \centering 
        \includegraphics[width=\textwidth]{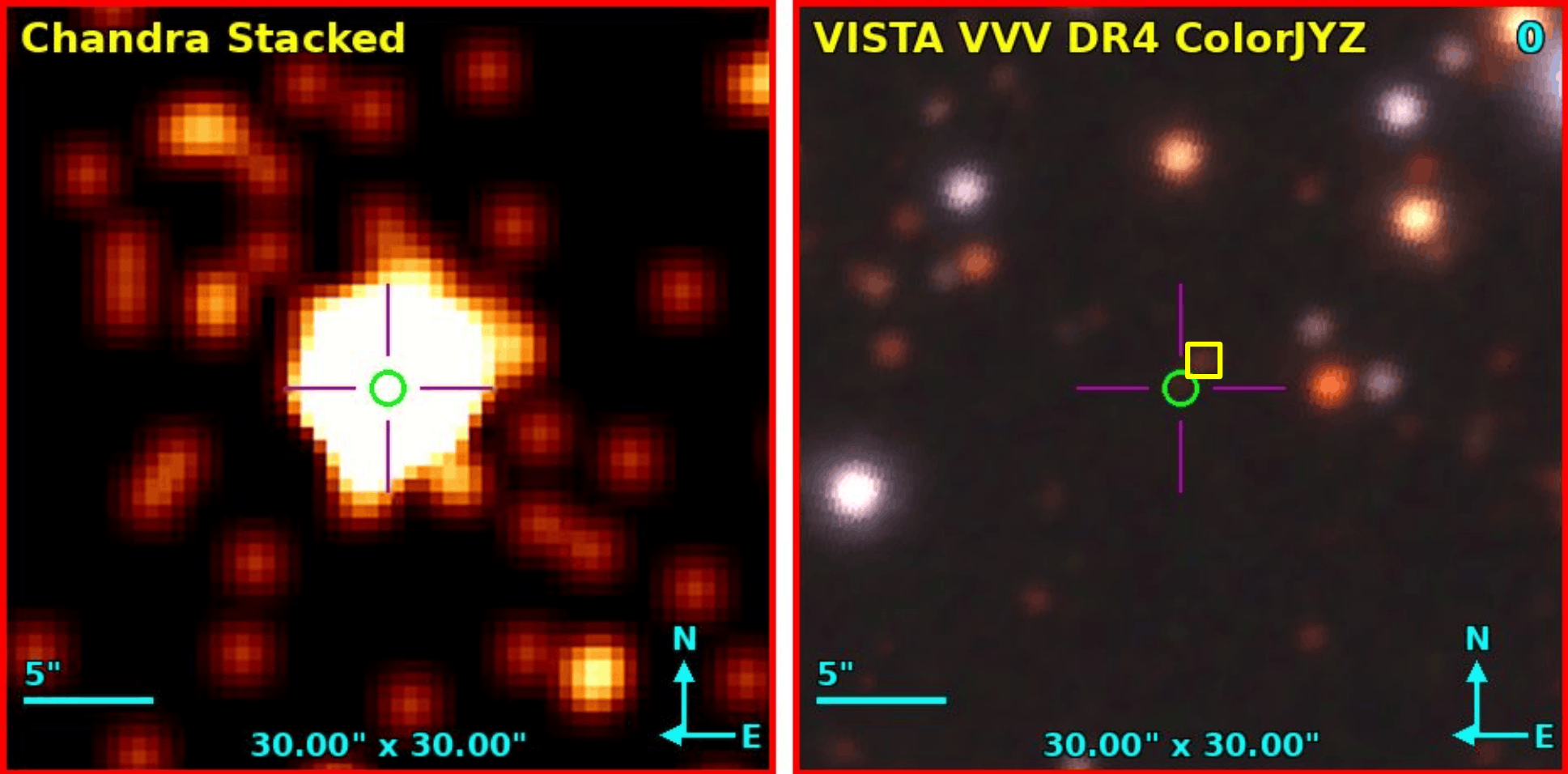}
        \caption{Src1} 
        \label{fig:Src1}
    \end{subfigure}
    \begin{subfigure}[b]{0.45\textwidth}
        \centering 
        \includegraphics[width=\textwidth]{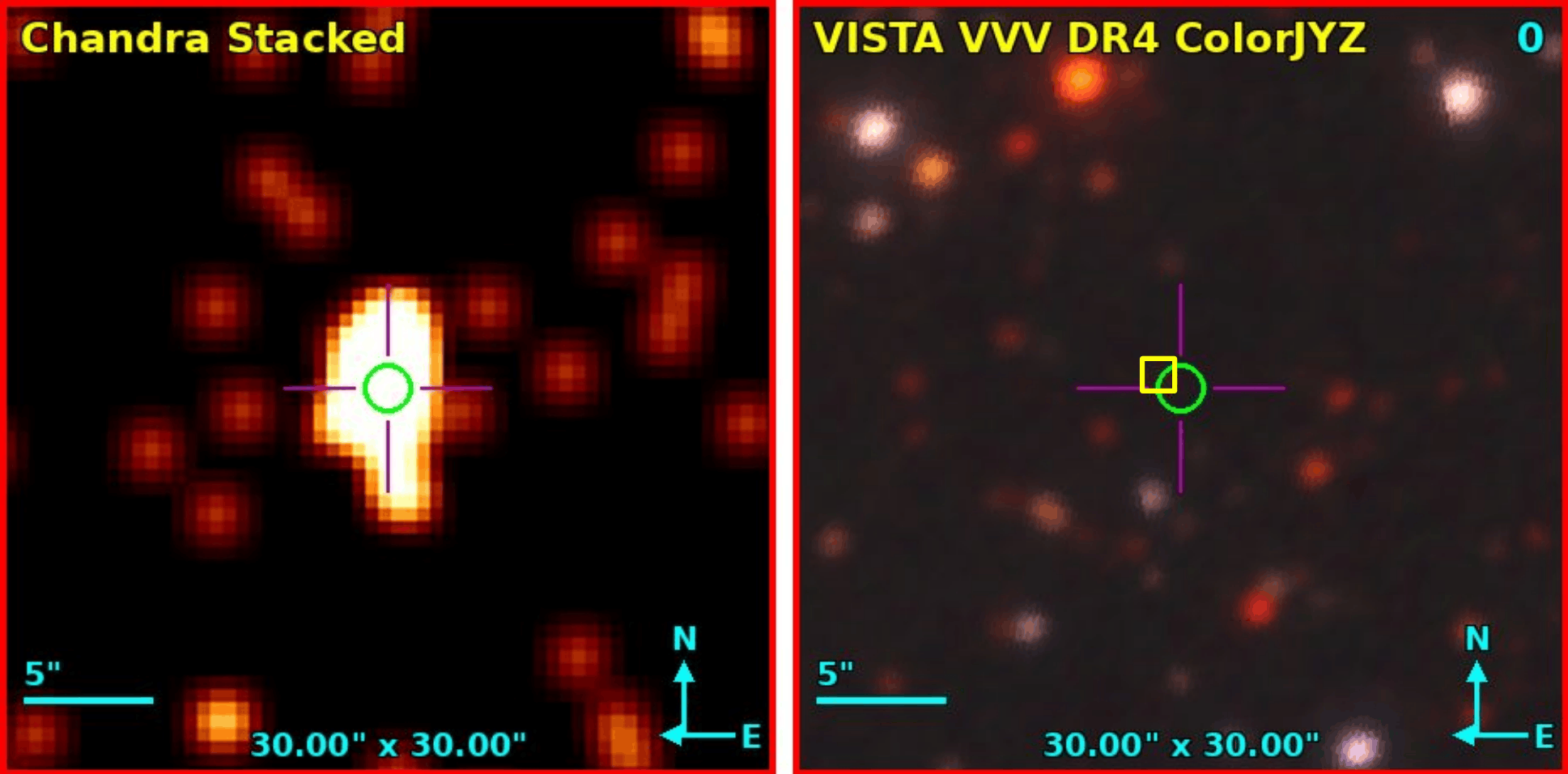}
        \caption{Src5} 
        \label{fig:Src5}
    \end{subfigure}
    \begin{subfigure}[b]{0.45\textwidth}
        \centering 
        \includegraphics[width=\textwidth]{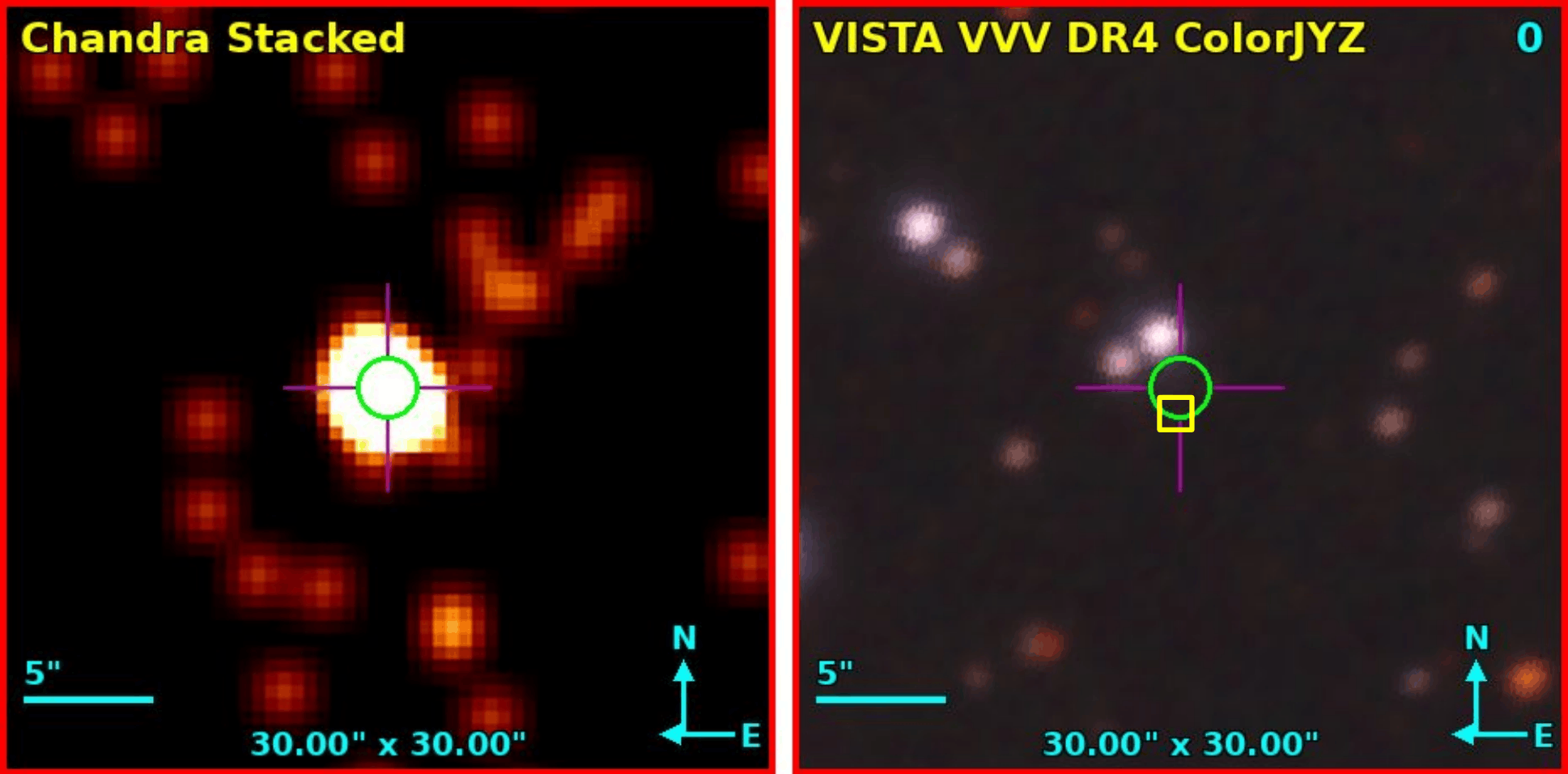}
        \caption{Src6} 
        \label{fig:Src6}
    \end{subfigure}
    \begin{subfigure}[b]{0.45\textwidth}
        \centering 
        \includegraphics[width=\textwidth]{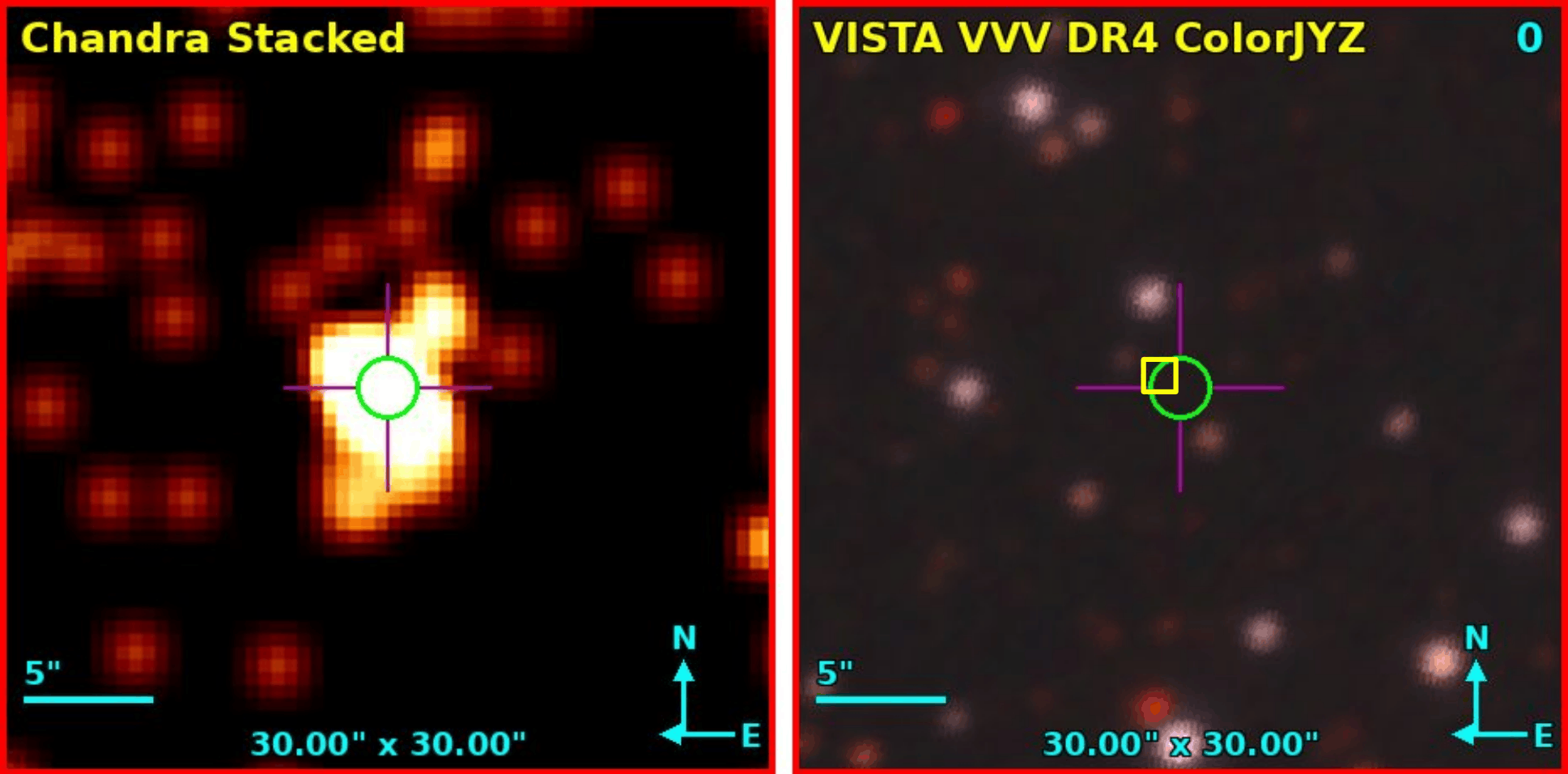}
        \caption{Src8} 
        \label{fig:Src8}
    \end{subfigure}
    \begin{subfigure}[b]{0.45\textwidth}
        \centering 
        \includegraphics[width=\textwidth]{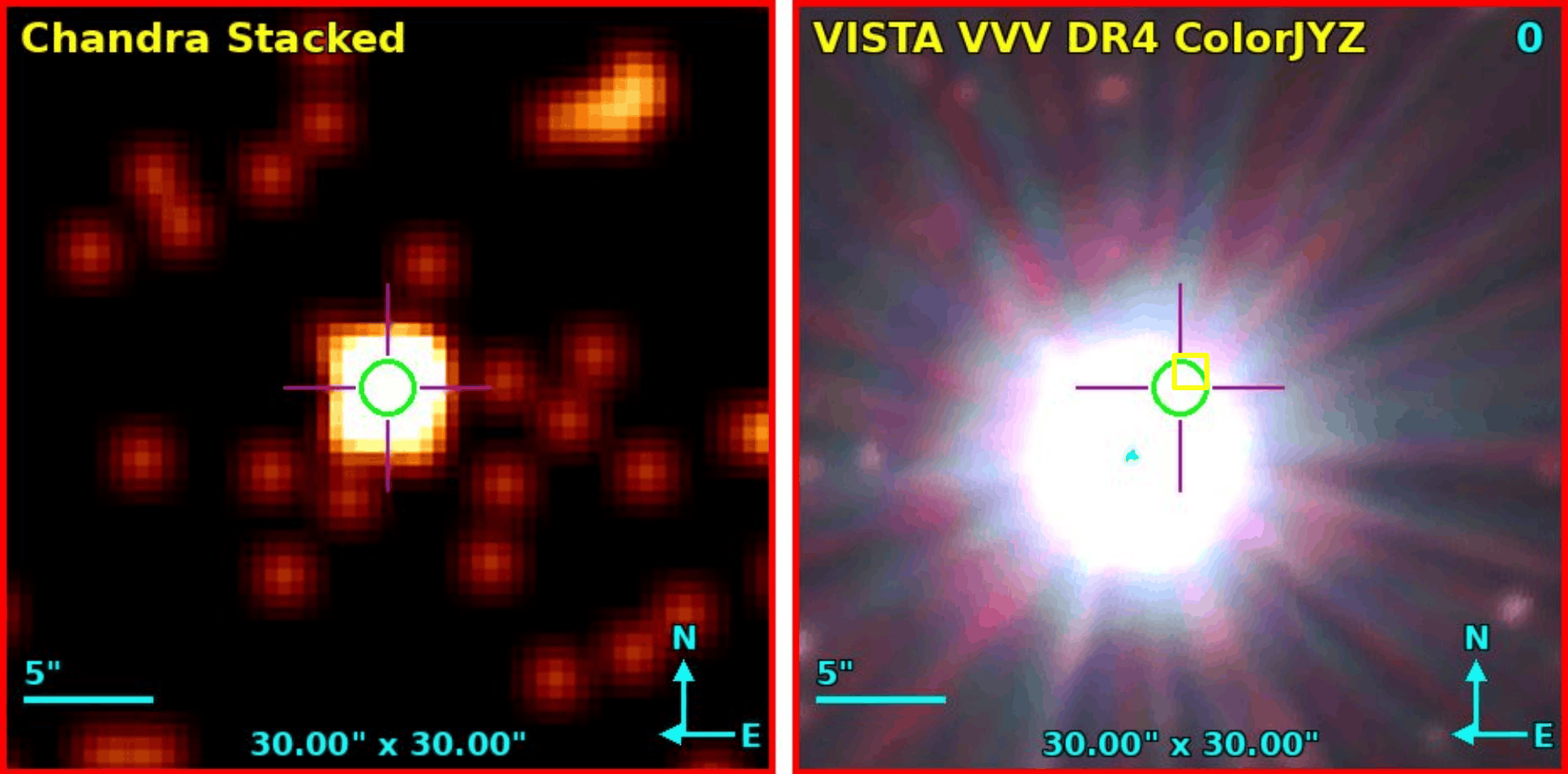}
        \caption{Src9} 
        \label{fig:Src9}
    \end{subfigure}
    \begin{subfigure}[b]{0.45\textwidth}
        \centering 
        \includegraphics[width=\textwidth]{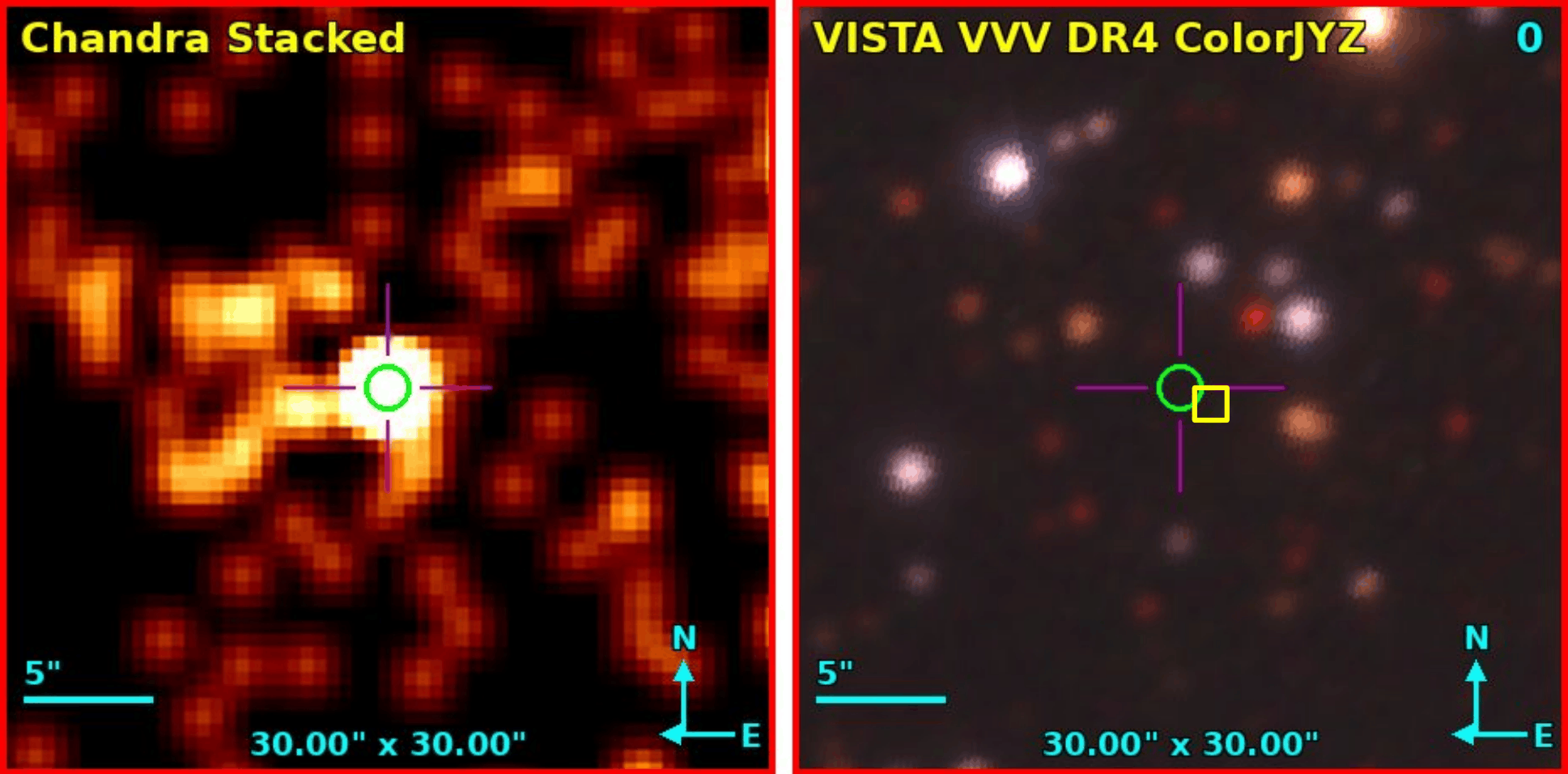}
        \caption{Src11} 
        \label{fig:Src11}
    \end{subfigure}
    \begin{subfigure}[b]{0.45\textwidth}
        \centering 
        \includegraphics[width=\textwidth]{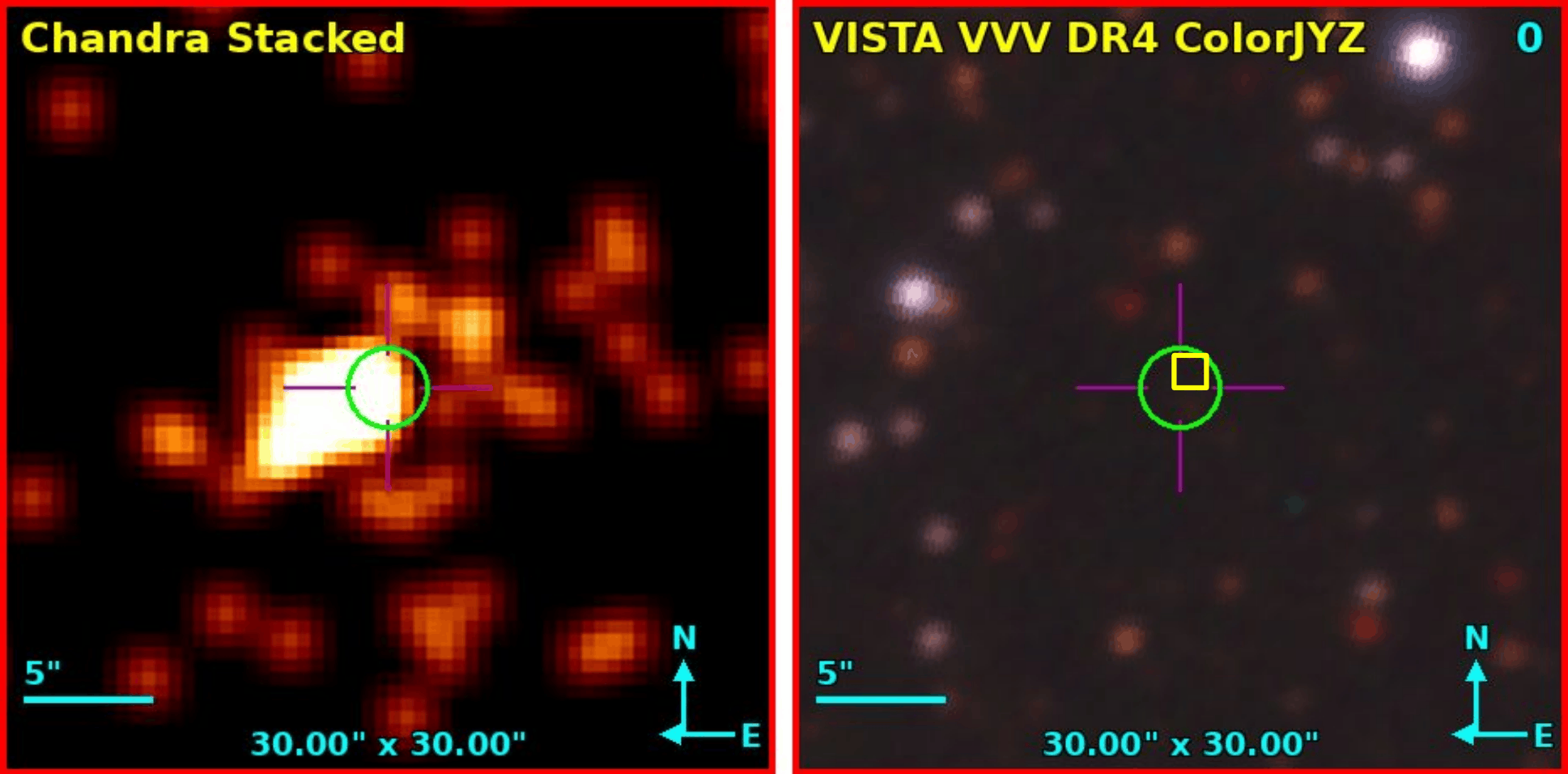}
        \caption{Src12} 
        \label{fig:Src12}
    \end{subfigure}
    \begin{subfigure}[b]{0.45\textwidth}
        \centering 
        \includegraphics[width=\textwidth]{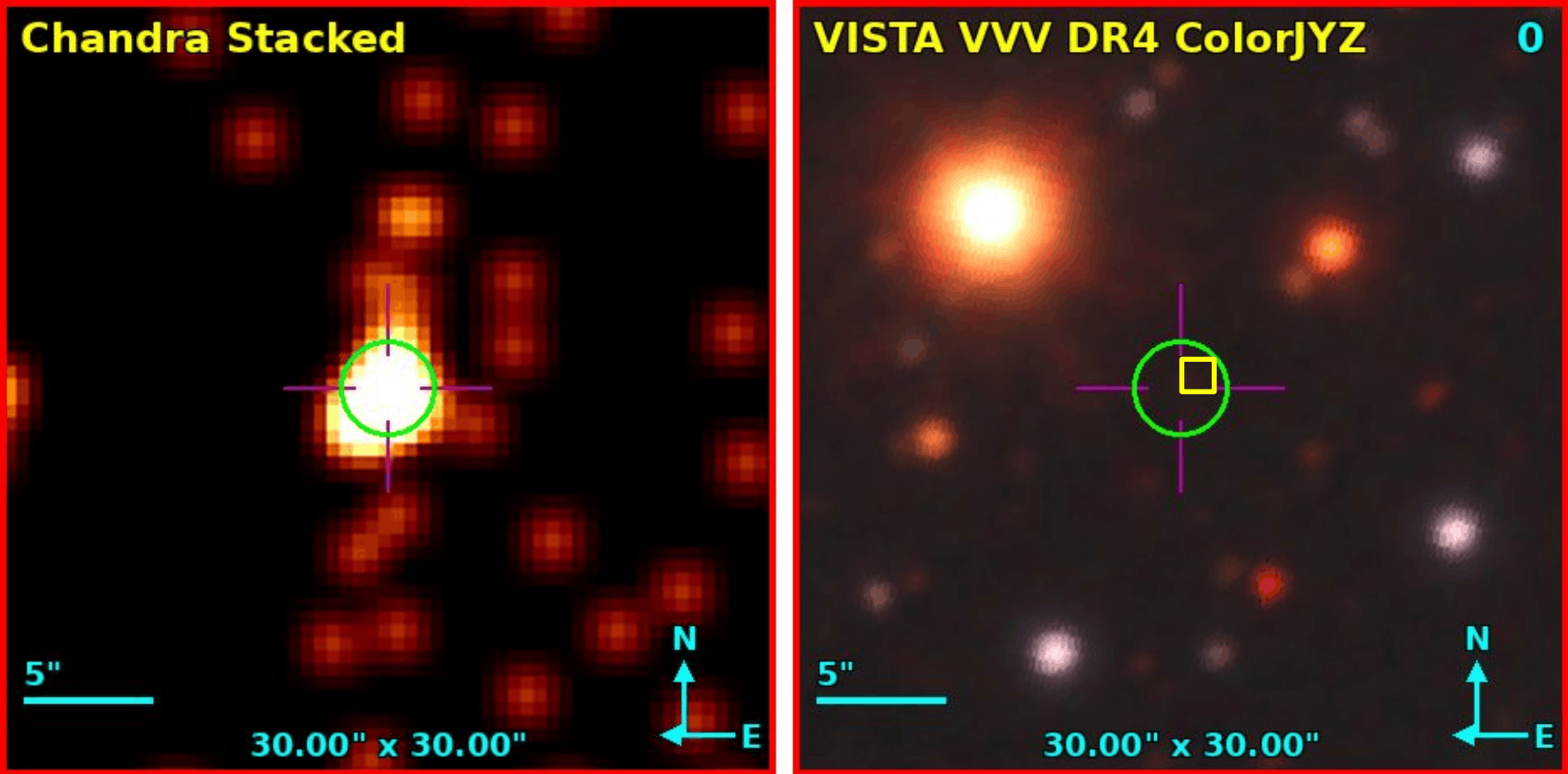}
        \caption{Src14} 
        \label{fig:Src14}
    \end{subfigure}
    \begin{subfigure}[b]{0.45\textwidth}
        \centering 
        \includegraphics[width=\textwidth]{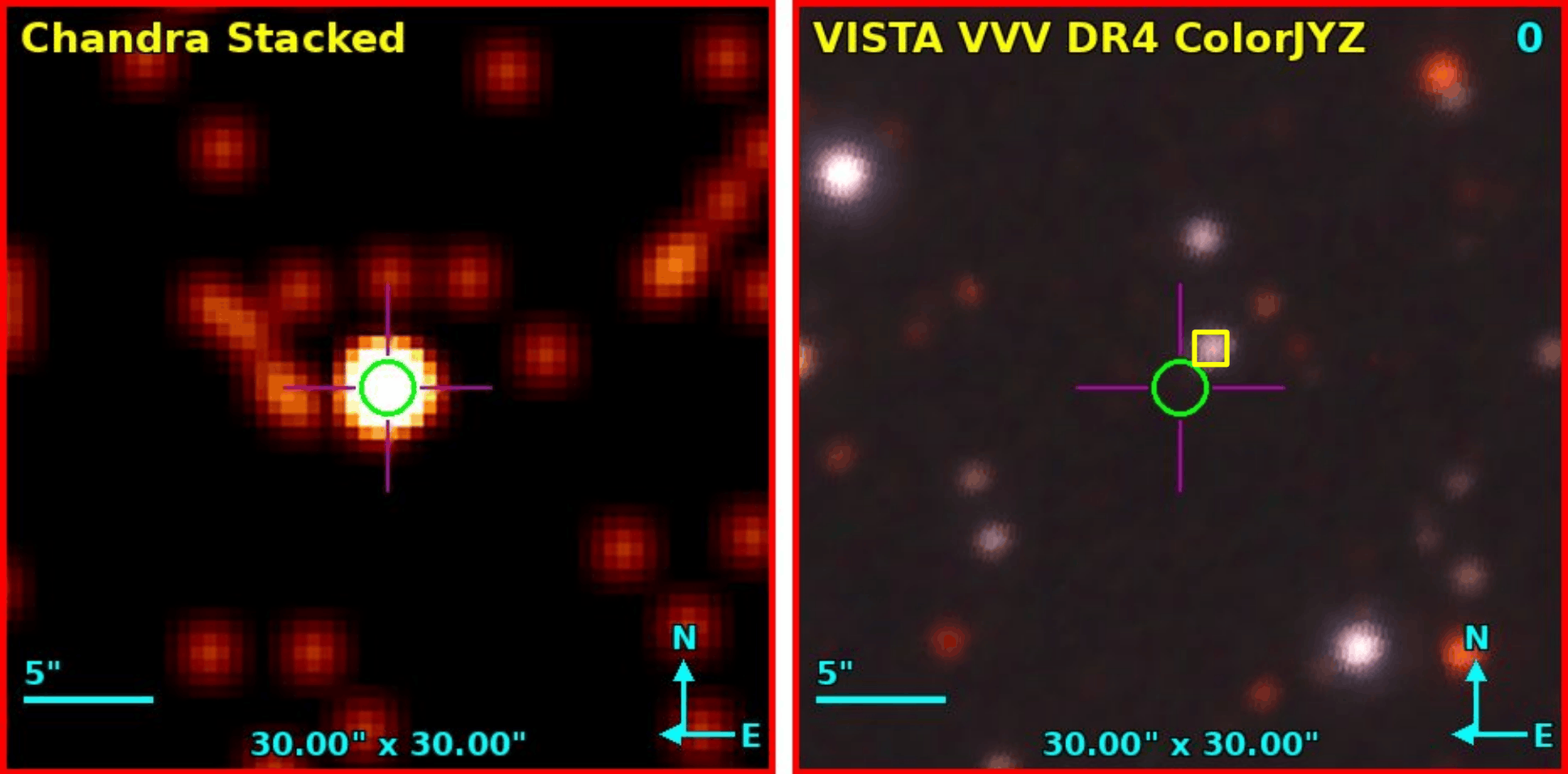}
        \caption{Src15} 
        \label{fig:Src15}
    \end{subfigure}

    \caption{MW images of the nine sources classified as NS candidates. For each source, the left panel shows the stacked \textit{Chandra} image smoothed with a Gaussian kernel of $4$ pixels, and  the right panel shows a VISTA VVV DR4 color $JYZ$ image. The X-ray positional error of 95\% confidence are shown in green circles and the closest NIR counterpart are shown in yellow squares. All panels have a field of view of $30''\times30''$.}
    \label{fig:MW}
\end{figure*}

The multiwavelength machine-learning classification (MUWCLASS) pipeline developed by \cite{2022ApJ...941..104Y} was used to perform an automated classification of 18 X-ray sources detected in the field of HESS J1702--420A. The pipeline trains a random forest machine-learning (ML) model with a training data set (TD) of $\sim3000$ sources of literature-verified classes from the Chandra Source Catalog version 2.0 (CSCv2.0), presented in \cite{2022ApJ...941..104Y} (with updates described in \citealt{2024ApJ...971..180Y}), including AGNs, low-mass stars (LM-STARs), high-mass stars (HM-STARs), young stellar objects (YSOs), CVs, high-mass X-ray binaries (HMXBs), low-mass X-ray binaries (LMXBs)\footnote{This class also includes nonaccreting X-ray binaries, such as wide-orbit binaries with millisecond pulsars (MSPs), as well as redback and black widow systems.}, and neutron stars (NSs). In addition to X-ray properties (fluxes, hardness ratios, and X-ray variability measures), each source is characterized by its 
optical/NIR/IR properties from the Gaia DR3 \citep{2023A&A...674A...1G}, 2MASS \citep{2006AJ....131.1163S}, AllWISE \citep{2014yCat.2328....0C}, and CatWISE2020 \citep{2021ApJS..253....8M} catalogs using the probabilistic cross-matching algorithm NWAY (\citealt{2018MNRAS.473.4937S}). 
The NWAY algorithm provides $p\_{\rm any}$\footnote{We kept the original notation from \cite{2018MNRAS.473.4937S} here.}, the probability of having a true counterpart among all associations considered, and $p\_{\rm i}$, the probability of each ($i$th) counterpart (including the case of having no association at all) being the right counterpart. 

Here, we discuss the automated classification results together with MW properties and variability, with particular emphasis on compact-object candidates and their potential contribution to the observed TeV emission. 
A summary of these results is available at \href{https://huiyang-astro.github.io/HESS_J1702-420/}{https://huiyang-astro.github.io/HESS\_J1702-420/}, which also provides an interactive interface for exploring the MW information, spectral fitting, and classification results of individual sources.
For further consideration, we selected sources classified as NS candidates that do not have an identified MW counterpart.

Besides the all sky surveys used by the MUWCLASS pipeline, we also retrieved NIR images centered on the source positions from the VISTA Variables in the Via Lactea survey \citep[VVV;][]{2010NewA...15..433M}. 
The VVV survey provides deep NIR imaging covering wavelengths from 0.9 to 2.5~$\mu$m, with limiting magnitudes ranging from 21.9 to 18.1~mag and a mean angular resolution of $\sim0.34''$, substantially improving upon the 2MASS survey, which has nominal point-source limits of $J=15.8$, $H=15.1$, and $K_s=14.3$~mag and an angular resolution of $\sim2.5''$. 
The high sensitivity and angular resolution of these NIR surveys are critical for identifying sources in the heavily absorbed Galactic plane, including flaring stars and obscured objects.

Figure~\ref{fig:MW} shows a subset of the sources listed in Table~\ref{tab:classification} that are 
classified as NS candidates by MUWCLASS and have no reliable counterpart identified through the all-sky MW catalogs used by the pipeline. 
For each source, the left panel shows the stacked \textit{Chandra} image, while the right panels show the corresponding VVV images, respectively.

\begin{figure*}
    \centering
    \begin{subfigure}[b]{0.82\textwidth} 
        \centering
        \includegraphics[width=\textwidth]{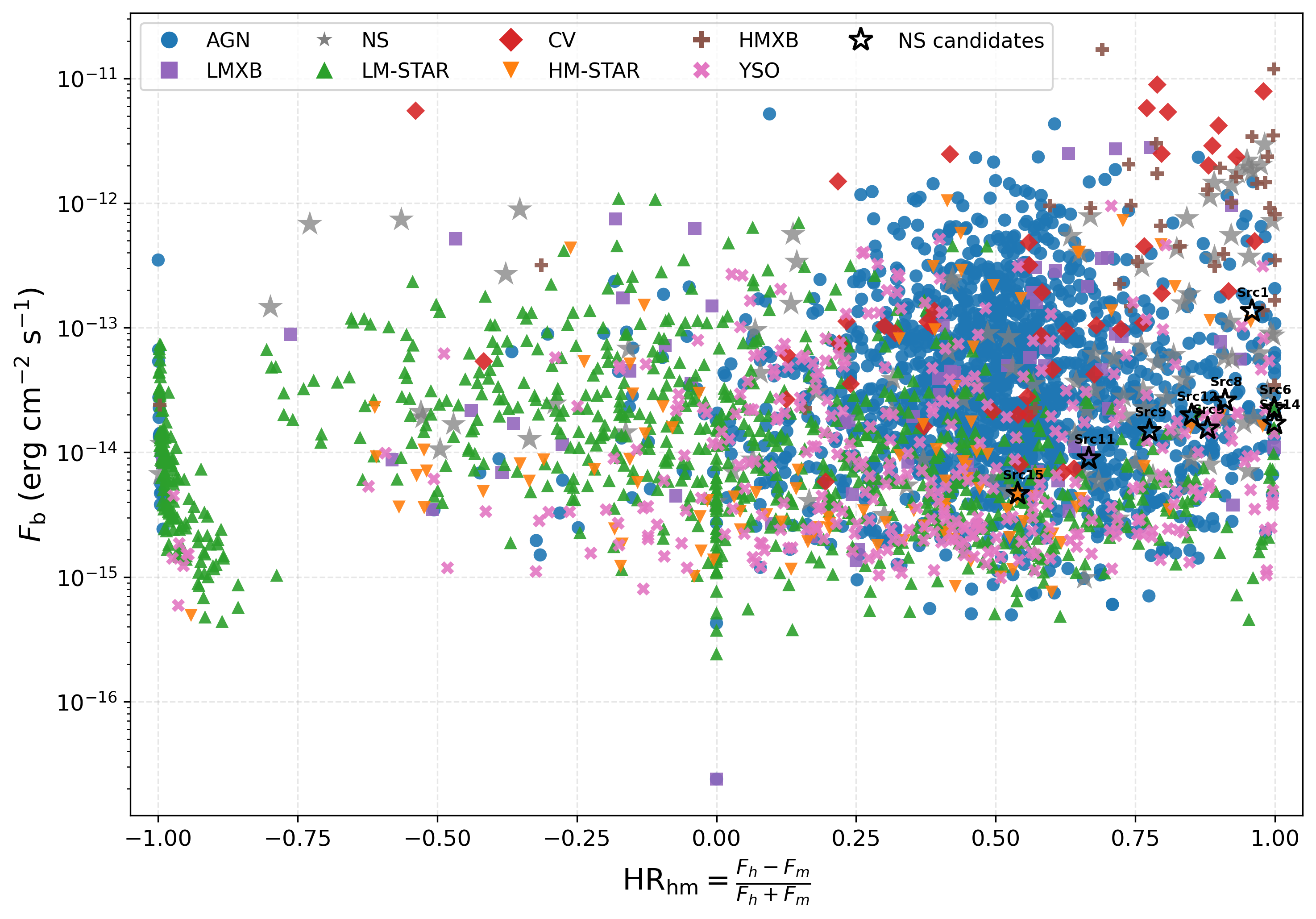}
        \label{fig:Fb_HR}
    \end{subfigure}
    \hfill
    \begin{subfigure}[b]{0.82\textwidth}  
        \centering 
        \includegraphics[width=\textwidth]{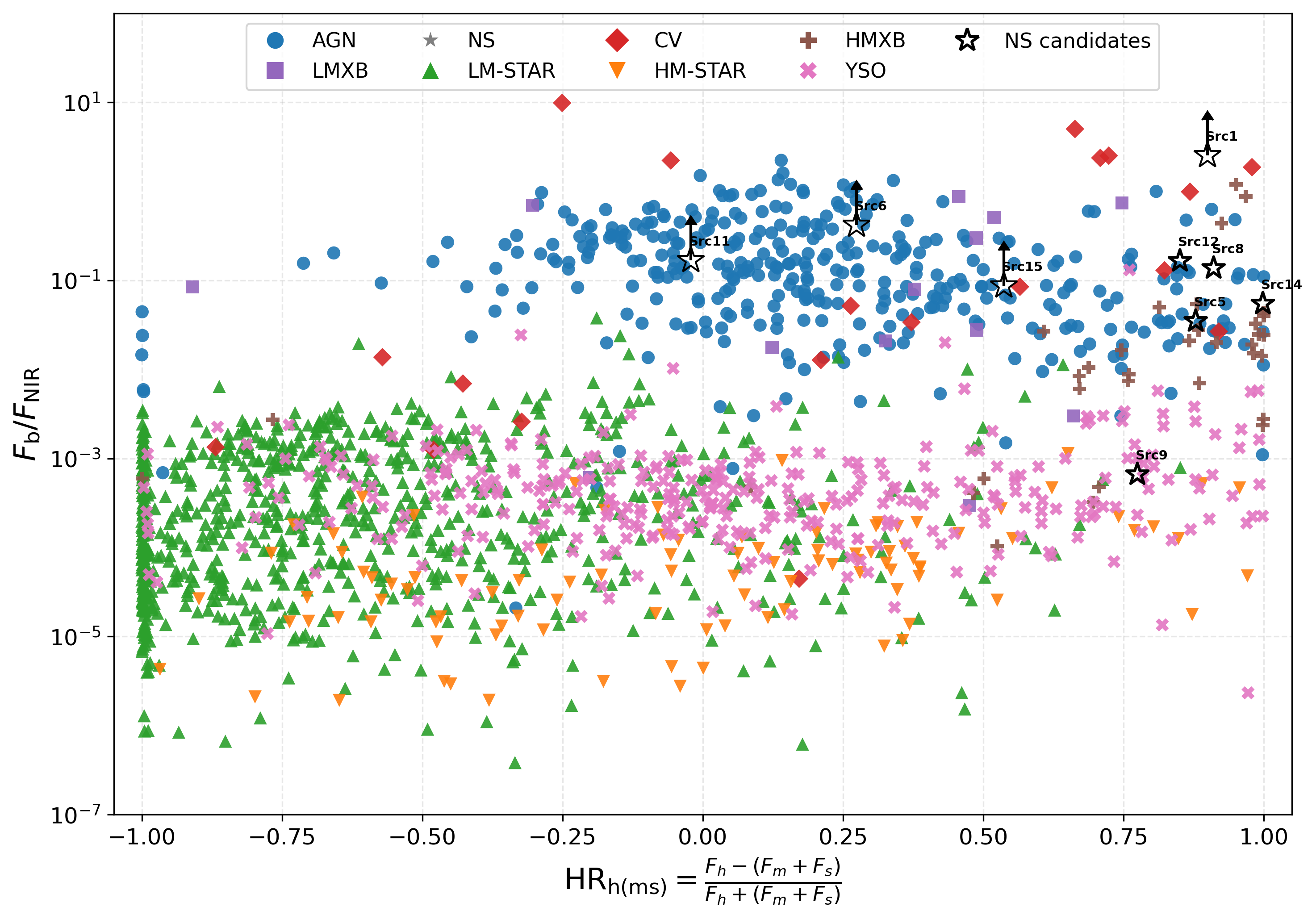}
        \label{fig:FX_FIR}
    \end{subfigure}
    \vskip\baselineskip
    \caption{Examples of two-dimensional feature spaces populated by literature-verified sources from the TD available at \href{https://muwclass.github.io/XCLASS_CSCv2/}{https://muwclass.github.io/XCLASS\_CSCv2/}, shown with different colors and markers, together with the NS candidates identified in the field of HESS~J1702--420 as hollow black stars. The top panel shows the broadband X-ray flux as a function of the X-ray hardness ratio $\mathrm{HR}_{hm}$, while the bottom panel shows the X-ray-to-NIR flux ratio, $F_{\rm X}/F_{\rm NIR}$, as a function of $\mathrm{HR}_{h(ms)}$. For the NS candidates, potential NIR counterparts were searched for using the deeper VVV survey. For sources without a detected NIR counterpart, lower limits on $F_{\rm X}/F_{\rm NIR}$ were calculated by adopting the nominal $5\sigma$ VVV limiting magnitude of $K_s=18.1$ mag (Vega system).}
    \label{fig:MUWCLASS-2D}
\end{figure*}

Figure~\ref{fig:MUWCLASS-2D} shows two examples of two-dimensional feature spaces populated by literature-verified sources in the TD and the NS candidates detected in the HESS~J1702--420A field. Different source classes occupy partially distinct regions and exhibit different clustering patterns in these feature spaces, which are among the characteristics exploited by the MUWCLASS classifier to distinguish between source populations (see. e.g.,  \citealt{2022ApJ...941..104Y}). The top panel shows the broadband X-ray flux, $F_{\rm b}$, as a function of the X-ray hardness ratio, $\mathrm{HR}_{hm}=(F_{\rm h}-F_{\rm m})/(F_{\rm h}+F_{\rm m})$. The bottom panel shows the X-ray-to-NIR flux ratio, $F_{\rm b}/F_{\rm NIR}$, as a function of $\mathrm{HR}_{h(ms)}=(F_{\rm h}-F_{\rm m}-F_{\rm s})/(F_{\rm h}+F_{\rm m}+F_{\rm s})$. Here, $F_{\rm b}$, $F_{\rm h}$, $F_{\rm m}$ and $F_{\rm s}$ denote the X-ray fluxes in the broad, hard, medium, and soft bands, respectively, as defined in Section~\ref{sec:obs_reduction}. 
The NIR flux, $F_{\rm NIR}$, is derived from the $K$-band magnitudes of the TD sources and the $K_s$-band magnitudes of the classified sources using the 2MASS and VVV surveys, respectively, following
$F_{\rm NIR}=9.3\times10^{-7}\times10^{-0.4K}\ {\rm erg\,cm^{-2}\,s^{-1}}$,
where $K$ denotes the corresponding $K$ or $K_s$ magnitude. 
For the NS candidates in the HESS~J1702--420A field, we searched for NIR counterparts using the deeper VVV survey. For sources without detected NIR counterparts, lower limits on $F_{\rm b}/F_{\rm NIR}$ were estimated by adopting the nominal $5\sigma$ VVV limiting magnitude of $K_s=18.1$~mag in the Vega system. The NS candidates are highlighted by hollow black stars.

Src11 appears to be the most plausible pulsar candidate because it is located close to the likely non-thermal extended emission, namely the eastern and western lobes (see Section~\ref{sec:cxo_image}). The source is classified by MUWCLASS as an NS with a probability of 66\% and a classification confidence threshold (CT, see its definition at equation (7) in \citep{2022ApJ...941..104Y})$=1.20$. It lacks any MW counterpart and shows no significant variability. Although its spectrum is poorly constrained, the photon index is compatible with a non-thermal pulsar spectrum when $N_{\rm H}$ is fixed to the value measured for the extended emission. Three NIR sources from the VISTA VVV ZYJHKs Catalog Data Release 4.2 \citep[hereafter VVV Catalog;][]{2012A&A...537A.107S}, namely vvv~J170144.39-421416.05, vvv~J170144.57-421416.84, and vvv~J170144.59-421413.70, are found at separations of $1.5''$, $1.6''$, and $1.9''$, respectively, all outside the 95\% positional uncertainty of $0.85''$ and therefore are not considered convincing counterparts. The nearby radio features and their possible interpretations are discussed in Section~\ref{sec:pulsar_PWN}. 

Although Src1 is the brightest in the field and is closer to HESS J1702-420A, the earlier analysis of  XMM-Newton  spectra revealed the detection of an Fe spectral line and the  measured high X-ray absorption column density (\mbox{N$_\text{H}=(5.1\pm 1.1)\times 10^{22}\,\text{cm}^{-2}$}), strongly favoring a CV classification \citepalias{2022A&A...667A.130G}. The spectral analysis with CXO data does not reveal such a feature, likely due to the smaller effective area of CXO. The source also appears to vary in flux between XMM-Newton, \Suzaku, and Chandra observations.  
However a periodic pulsation search in EPIC data results in a negative outcome (see Section \ref{sec:variability}). 
There is a faint NIR counterpart vvv J170157.09-421025.62 1.2 arcsec away from the VVV Catalog, 
outside the 95\% PU of 0.67$''$.   
Since the MUWCLASS pipeline does not use cross-mission variability and X-ray spectral lines as a feature and the possible NIR counterpart is too dim to be detected in all sky surveys currently used by MUWCLASS, it classified the source as a NS with fairly high confidence. 

Src15 is classified as an NS with a high CT$=2.45$ and no MW counterpart has found in deeper surveys within its 95\% PU. However, the limited number of counts makes the spectral and temporal properties of Src15 not well constrained, making it difficult to investigate its nature in detail. However, the point-like morphology of the source (see Fig \ref{fig:MW}) conflicts the scenario of a NS in a PWN.

Src5 and Src8 both have relatively high hardness ratios $HR_{\rm h(ms)}$ of 0.88 and 0.91, and are not statistically significantly variable. Hence, they are classified as a NS with high probabilities of 84\% (CT$=3.30$) and 84\% (CT$=3.00$).  They are located at about the same offset from HESS J1702-420A as Src11 but they do not have any diffuse emission around them. 
However, each of them has a potential NIR counterpart vvv J170213.19-421456.24 
at a separation of $0.6''$ for Src5, and vvv J170220.11-421536.43 	
at a separation of $0.94''$ for Src8, from the VVV Catalog, all within their 95\% PU. 

For Src6, Src9, Src12, and Src14, they are classified as a NS with a low to modest CT from 0.51 to 1.15, as no counterpart are found using the all-sky survey used by MUWCLASS. However, all of them are found with potential NIR counterparts within their 95\% PU from the VVV Catalog: one NIR counterpart vvv J170131.61-421200.30 
at a separation of 1.2$''$ is found for Src6; one NIR counterpart vvv J170203.87-421736.78 
at a separation of 0.3$''$ is found for Src9; two NIR counterparts vvv J170217.13-420956.15 
at a separation of 0.7$''$ and vvv J170217.12-420957.81 
at a separation of 1.1$''$ are found for Src12; one NIR counterpart vvv J170219.61-421411.93  
at a separation of 0.3$''$ is found for Src14. For Src6, its relatively soft spectrum ($\Gamma=1.8^{+1.4}_{-1.2}$) and significant variability suggests coronal flare in a star/YSO or active binary.

\section{Conclusions}
\label{sec:conclusion}

We present a comprehensive \textit{Chandra} study of the field of the PeVatron candidate HESS~J1702--420A, motivated by the previously reported hard and apparently extended X-ray emission from XMMU~J170147.3--421407 by \citetalias{2022A&A...667A.130G}. The new \textit{Chandra} observations confirm the complex and extended nature of this emission and resolve XMMU~J170147.3--421407 into a faint point-like source, Src11 or CXO J170144.5--421415, and asymmetric diffuse emission that can be separated into distinct eastern and western lobes.

The spectral and morphological properties of the Src11$+$extended-emission system make it a plausible pulsar/PWN candidate. The spectra of Src11 and the diffuse emission are adequately described by absorbed power-law models, and the measured photon indices are consistent with the range commonly observed in pulsar/PWN systems. 
Furthermore, assuming a distance of 4.7\,kpc, both the relation between the X-ray luminosities of Src11 and the diffuse emission and the relation between the X-ray luminosity of the diffuse emission and the $\gamma$-ray luminosity of HESS~J1702--420A are consistent with the distributions of known PWNe in the $L_{\rm X,psr}$--$L_{\rm X,pwn}$ and $L_{\rm X}$--$L_{\gamma}$ planes, respectively. 
Src11 is also classified as a NS candidate by the MUWCLASS pipeline and has no convincing MW counterpart. Although the model-independent variability analysis initially suggest possible inter-observation variability, the model-dependent fluxes derived from the spectral fits are statistically consistent within their uncertainties and do not provide compelling evidence for significant long-term variability. Together with the absence of significant short-term variability, these properties are more naturally consistent with an isolated pulsar than with an accreting compact object and make Src11 the most plausible pulsar candidate physically associated with the diffuse X-ray emission.

The morphology of the extended emission, however, does not readily correspond to a conventional PWN geometry, such as a torus/jet or pulsar-tail structure. In particular, the eastern and western lobes are not suggestive of a high-Mach-number bow shock, as would be expected if the $\sim7'$ offset between Src11 and the center of HESS~J1702--420A were produced by pulsar motion over several tens of kyr.  
Alternatively, if the putative pulsar remains within its host SNR, the eastern and western lobes could potentially represent a jet and counterjet from a subsonic PWN whose morphology has been distorted by interactions with the surrounding SNR environment. In this scenario, the offset TeV source could represent a relic PWN displaced through interactions with the SNR reverse shock.

We also note the presence of a possible shell-like radio structure northwest of Src11 in the RACS low-frequency image. Its nature remains uncertain, but it could represent a radio-bright part of a relic PWN physically associated with Src11 and, potentially, with the HESS~J1702--420A region.

The MUWCLASS results and deeper MW imaging also provide important constraints on the other compact-object candidates in the field. Several sources classified as NS candidates are either associated with possible NIR counterparts detected in deeper surveys and/or have X-ray properties inconsistent with a young pulsar/PWN interpretation. 
Taken together, these results make Src11 the most compelling pulsar/PWN candidate among the sources detected in the present \textit{Chandra} observations.

Overall, the new \textit{Chandra} observations strengthen the case that XMMU~J170147.3--421407 contains a compact-object/PWN system and identify Src11/CXO~J170144.5--421415 as the most plausible pulsar candidate currently known in the vicinity of HESS~J1702--420A. However, the complex and asymmetric morphology of the diffuse emission and the offset between Src11 and the TeV source prevent a firm physical association. A relic-PWN interpretation remains possible but requires an unusual evolutionary history and does not straightforwardly explain the observed X-ray morphology. The possible shell-like radio structure further suggests that the region may contain a more complex SNR/PWN environment, leaving open alternative scenarios for the origin of the TeV emission. 

Further MW observations will be important for resolving the nature of this system. Deeper X-ray observations could better constrain the spectrum and variability of Src11 and search for spectral evolution across the diffuse emission. In particular, long-baseline X-ray observations could enable a measurement of the proper motion of Src11, providing an important test of its pulsar nature and of its possible physical association with the extended X-ray and TeV emission. More sensitive and higher-angular-resolution radio observations could establish the nature of the shell-like structure and search for radio emission from a pulsar or PWN. Improved TeV measurements of the morphology and spectrum of HESS~J1702--420A will also be crucial for determining whether the TeV emission is physically associated with the Src11/PWN system, an SNR environment, or a distinct particle accelerator projected within the same complex Galactic region.

\begin{acknowledgements}
Support for this work was provided by the National Aeronautics and Space Administration through Chandra Award Number GO4-25043X issued by the Chandra X-ray Center, which is operated by the Smithsonian Astrophysical Observatory for and on behalf of the National Aeronautics Space Administration under contract NAS8-03060. 

This paper employs a list of Chandra datasets, obtained by the Chandra X-ray Observatory, contained in the Chandra Data Collection (CDC). 
\end{acknowledgements}

%

\bibliographystyle{bibtex/aa} 
\bibliography{biblio.bib} 

\onecolumn

\begin{sidewaystable*}
\caption{MW properties and MUWCLASS Classification results for CSC sources (S/N$>5$) in the field.}
\label{tab:classification}
\centering
\begin{tabular}{lllllllllllllllll}
\hline        
Source &  Class$^a$ & Class prob.$^b$ & CT & RA$^c$ & Dec$^c$ & PU$^d$  & Sig.$^e$ & $F_{\rm b}$$^f$ & $HR_{h(ms)}$ &  $P_{\rm intra}$ &  $P_{\rm inter}$ & $p_{\rm any}$ & $p_i$ &   $G$mag$^g$ & $J$mag$^h$ & $W1$mag$^i$ \\
 &  &  & & deg & deg & $''$ (95\%) &  & $10^{-14}$\,cgs &  &  &  &  &  &  mag & mag & mag \\
\hline
\textbf{Src1} & NS & 0.75 & 2.30 & 255.48822 & -42.17403 & 0.67 & 48.1 & 13.6 & 0.90 & 0.963 & 0.881 & 0.00 & 0.00 &  &  &  \\
Src2 & NS & 0.42  & 0.03 & 255.50154 & -42.22118 & 0.67 & 27.7 & 0.9 & 0.18 & 0.877 & 1.000 & 0.99 & 0.00 &  &  &  \\
 & YSO & 0.61 & 2.30 & 255.50154 & -42.22118 & 0.67 & 27.7 & 0.9 & 0.18 & 0.877 & 1.000 & 0.99 & 1.00 & 19.97 & 14.01 &  \\
Src3 & NS & 0.48  & 1.03 & 255.32461 & -42.22716 & 1.18 & 16.3 & 5.0 & 1.00 & 0.830 & 1.000 & 0.91 & 0.00 &  &  &  \\
 & YSO & 0.42  & 1.77 & 255.32461 & -42.22716 & 1.18 & 16.3 & 5.0 & 1.00 & 0.830 & 1.000 & 0.91 & 1.00 &  &  & 13.65 \\
Src4 & LMXB & 0.58  & 1.08 & 255.40013 & -42.32877 & 1.42 & 12.8 & 1.5 & 0.39 & 0.982 & 0.970 & 0.97 & 0.00 &  &  &  \\
 & HM-STAR & 0.43  & 0.53 & 255.40013 & -42.32877 & 1.42 & 12.8 & 1.5 & 0.39 & 0.982 & 0.970 & 0.97 & 1.00 & 15.65 & 11.20 & 9.15 \\
Src5 & NS & 0.84 & 3.30 & 255.55473 & -42.24901 & 0.93 & 11.9 & 1.6 & 0.88 & 0.169 & 0.514 & 0.00 & 0.00 &  &  &  \\
Src6 & NS & 0.61  & 1.15 & 255.38165 & -42.19976 & 1.19 & 11.6 & 2.2 & 0.27 & 0.892 & 0.962 & 0.00 & 0.00 &  &  &  \\
Src7 & LMXB & 0.57 & 0.46 & 255.42984 & -42.24404 & 0.86 & 9.5 & 0.7 & -0.69 & 0.877 & 0.992 & 0.99 & 0.00 &  &  &  \\
 & YSO & 0.54 & 0.83 & 255.42984 & -42.24404 & 0.86 & 9.5 & 0.7 & -0.69 & 0.877 & 0.992 & 0.99 & 1.00 & 15.90 & 13.12 &  \\
Src8 & NS & 0.84  & 3.00 & 255.58347 & -42.26023 & 1.19 & 9.2 & 2.6 & 0.91 & 0.288 & 0.273 & 0.00 & 0.00 &  &  &  \\
Src9 & NS & 0.56  & 0.51 & 255.51620 & -42.29363 & 1.06 & 9.1 & 1.5 & 0.77 & 0.787 & 0.230 & 0.00 & 0.00 &  &  &  \\
Src10 & NS & 0.74  & 1.91 & 255.49810 & -42.18944 & 1.05 & 8.7 & 1.2 & -0.29 & 0.403 & 0.991 & 0.99 & 0.00 &  &  &  \\
 & YSO & 0.90  & 6.03 & 255.49810 & -42.18944 & 1.05 & 8.7 & 1.2 & -0.29 & 0.403 & 0.991 & 0.99 & 0.51 & 16.70 & 12.48 &  \\
 & YSO & 0.36 & 0.56 & 255.49810 & -42.18944 & 1.05 & 8.7 & 1.2 & -0.29 & 0.403 & 0.991 & 0.99 & 0.49 &  &  & 10.12 \\
\textbf{Src11} & NS & 0.66  & 1.20 & 255.43547 & -42.23760 & 0.85 & 7.9 & 0.9 & -0.02 & 0.057 & 0.966 & 0.00 & 0.00 &  &  &  \\
Src12 & NS & 0.62  & 0.73 & 255.57152 & -42.16578 & 1.60 & 6.7 & 2.0 & 0.85 & 0.830 & 0.411 & 0.00 & 0.00 &  &  &  \\
Src13 & NS & 0.65  & 1.06 & 255.53649 & -42.20000 & 1.25 & 6.4 & 4.6 & -0.84 & 0.755 & 0.708 & 0.97 & 0.00 &  &  &  \\
 & YSO & 0.31  & 0.13 & 255.53649 & -42.20000 & 1.25 & 6.4 & 4.6 & -0.84 & 0.755 & 0.708 & 0.97 & 1.00 & 17.45 & 14.68 &  \\
Src14 & NS & 0.57 & 1.10 & 255.58181 & -42.23667 & 1.89 & 6.2 & 1.7 & 1.00 & 0.409 & 0.914 & 0.00 & 0.00 &  &  &  \\
\textbf{Src15} & NS & 0.75  & 2.45 & 255.46083 & -42.28065 & 1.08 & 5.8 & 0.5 & 0.54 & 0.135 & 0.893 & 0.00 & 0.00 &  &  &  \\
Src16 & NS & 0.85  & 3.53 & 255.46114 & -42.39037 & 3.66 & 5.6 & 1.5 & 0.90 & 0.042 & 0.788 & 0.68 & 0.00 &  &  &  \\
 & YSO & 0.47  & 1.84 & 255.46114 & -42.39037 & 3.66 & 5.6 & 1.5 & 0.90 & 0.042 & 0.788 & 0.68 & 0.51 &  &  & 14.43 \\
 & YSO & 0.49  & 1.86 & 255.46114 & -42.39037 & 3.66 & 5.6 & 1.5 & 0.90 & 0.042 & 0.788 & 0.68 & 0.49 &  &  & 14.89 \\
Src17 & LMXB & 0.56 & 0.96 & 255.49092 & -42.26319 & 1.04 & 5.4 & 0.3 & 0.16 & 0.999 & 0.980 & 0.98 & 0.00 &  &  &  \\
 & LMXB & 0.40  & 0.91 & 255.49092 & -42.26319 & 1.04 & 5.4 & 0.3 & 0.16 & 0.999 & 0.980 & 0.98 & 1.00 & 20.24 & 16.32 &  \\
Src18 & NS & 0.65  & 1.39 & 255.62670 & -42.30814 & 2.90 & 5.4 & 4.8 & 0.35 & 0.107 & 0.988 & 0.45 & 0.00 &  &  &  \\
 & CV & 0.44  & 0.73 & 255.62670 & -42.30814 & 2.90 & 5.4 & 4.8 & 0.35 & 0.107 & 0.988 & 0.45 & 1.00 & 17.81 & 15.18 &  \\
\hline
\end{tabular}
\tablefoot{
Source names in bold are NS candidates classified by MUWCLASS that have no MW counterpart within their 95\% PU, including the deeper VVV survey. \\
$^a$ MUWCLASS classification. \\
$^b$ Classification probability.  \\
$^c$ Right ascension and declination of the Chandra sources. \\
$^d$ The 2$\sigma$ X-ray positional uncertainty. \\
$^e$ X-ray detection significance. \\
$^f$ Average broadband flux in the 0.5--7\,keV band calculated with \textit{srcflux}. These values are model-independent and therefore differ from the model-dependent fluxes derived from the spectral fits in Table~\ref{tab:spectral_fit}.  \\
$^g$ Gaia G-band magnitude. \\
$^h$ 2MASS J-band magnitude.\\
$^i$ WISE W1-band magnitude. }
\end{sidewaystable*}

\begin{appendix}





\onecolumn
\section{Astrometry
\label{sec:astrometry}}

We co-aligned the X-ray images from individual Chandra observations by first registering all Chandra images to ObsID 30956, and then registering the merged ACIS image to the Gaia DR3 catalog \citep{2023A&A...674A...1G}, employing the same method used by \cite{2024ApJ...968...67G} (details are provided in the Appendix of that paper). 
As part of this procedure, we ran the CIAO tool \texttt{wavdetect} \citep{Freeman2002}  on all 3 observations detecting 15, 15, and 4 sources with significance $>3$ and net counts $>5$ in ObsIDs 28074, 30956 and 30957, respectively. We also ran {\tt wavdetect} on the merged ACIS-I image, detecting 24 sources with significance $>3$ and net counts $>10$. 
The CIAO tool \texttt{wcs\_match} is used to determine the translational shift that minimizes the root-mean-square residual (RMSR) between the matched pairs, achieving improvements of  11\%, 8\% and 20\% for ObsID=30957, 28074 (with respect to 30956), and the merged image, respectively. 
The 1$\sigma$ uncertainty of the astrometric correction is calculated using the weighted root-mean-square residual \citep[WRMSR, see equations A1 and A2 at ][]{2024ApJ...968...67G}. 
The WCS information and the aspect history files  are updated accordingly using \texttt{wcs\_update}.
The relative and absolute astrometry solutions are summarized in Table \ref{tab:astro_solution}.

\begin{table*}[ht!]
\caption{\label{tab:astro_solution} The Relative Astrometry Solutions in the First Part of the Table with the Reference ObsID $=30956$ and the Absolute Astrometry Solution (Corrected to Gaia DR3) in the Second Part of the Table}
\centering
\begin{tabular}{lcccccccc}
\hline\hline
ObsID & Radius$^a$ & Reslim$^b$ & 
$\Delta$R.A.$^c$  &
$\Delta$Decl.$^c$ & Prior RMSR  & Transformed RMSR  &
Transformed WRMSR& 
No. of pairs \\
 & ($\arcsec$) & ($\arcsec$) & 
($\arcsec$) &
($\arcsec$) &  ($\arcsec$) &  ($\arcsec$) &
($\arcsec$) & 
\\
\hline
30957 & 1.2 & 1.1  & -0.50  & -0.08 & 0.82 & 0.73 & 0.57 & 3 \\
28074 & 1.2& 1.0 & 0.05  & 0.29 & 0.75 & 0.69 & 0.69 & 7 \\
\hline 
Merged & 1.0 & 0.4  & 0.19 & 0.10 & 0.30 & 0.24 & 0.22 & 3 \\
\hline
\end{tabular}
\tablefoot{
$^a$ Source-matching radius used by \texttt{wcs\_match}. \\
$^b$ Residual limit used by \texttt{wcs\_match}. \\
$^c$ Astrometric corrections in right ascension and declination, respectively.
}
\end{table*}

\FloatBarrier 
\twocolumn

\end{appendix}
\end{document}